\documentclass[useAMS,usenatbib]{mn2e}
\usepackage{graphicx}
\usepackage{multirow}
\usepackage[perpage,para,symbol*]{footmisc}
\usepackage{siunitx}
\usepackage{natbib}

\defcitealias{Ercolano2010}{ECH11}

\title[Disc Clearing of Young Stellar Objects]{Disc Clearing of Young Stellar Objects: Evidence for Fast Inside-out Dispersal}
\author[Koepferl et al. 2012]
  {C. M.~Koepferl$^1$$^{\mbox{\thanks{koepferl@usm.lmu.de}}}$, B.~Ercolano$^{1,2}$, J.~Dale$^{1,2}$, P. S. Teixeira$^{3,4}$, T. Ratzka$^1$, L.~Spezzi$^5$\\
  $^1$Universit\"atssternwarte M\"unchen, Ludwig-Maximilians-Universit\"at, Scheinerstr. 1, 81679 M\"unchen, Germany\\
  $^2$Exzellenzcluster Universe, Boltzmannstr. 2, 85748 Garching bei M\"unchen, Germany\\
  $^3$Institut f\"ur Astrophysik, Universit\"at Wien, T\"urkenschanzstr. 17, 1180 Wien, Austria\\
  $^4$Laborat\'orio Associado Instituto D. Luiz-SIM, Universidade de Lisboa, Campo Grande, 1749-016, Lisboa, Portugal\\
  $^5$ESO Headquarters, Karl-Schwarzschild-Str. 2, 85748 Garching bei M\"unchen, Germany }
\date{}
\begin{document}

\pagerange{\pageref{firstpage}--\pageref{lastpage}} \pubyear{2012}

\maketitle

\label{firstpage}

\begin{abstract}
The time-scale over which and the modality by which young stellar objects (YSOs) disperse their circumstellar discs dramatically influences the eventual formation and evolution of planetary systems. By means of extensive radiative transfer (RT) modelling, we have developed a new set of diagnostic diagrams in the infrared colour-colour plane (K\,-\,[24] vs. K\,-\,[8]), to aid with the classification of the evolutionary stage of YSOs from photometric observations. Our diagrams allow the differentiation of sources with unevolved (primordial) discs from those evolving according to different clearing scenarios (e.g.\,homologous depletion vs. inside-out dispersal), as well as from sources that have already lost their disc. Classification of over 1500 sources in 15 nearby star-forming regions reveals that approximately 39\,\% of the sources lie in the primordial disc region, whereas between 31\,\%  and 32\,\% disperse from the inside-out and up to 22\,\% of the sources have already lost their disc. Less than 2\,\% of the objects in our sample lie in the homogeneous draining regime. Time-scales for the transition phase are estimated to be typically a few $10^{5}$\,years independent of stellar mass.
Therefore, regardless of spectral type, we conclude that currently available infrared photometric surveys point to fast (of order 10\,\% of the global disc lifetime) inside-out clearing as the preferred mode of disc dispersal.
\end{abstract}


\section{Introduction}
\label{Introduction}
Protoplanetary discs are an unavoidable consequence of the star formation process and harbour the material from which planets may form. For this reason, it is essential to understand their properties, evolution and final dispersal. For 
a recent review on protoplanetary discs and their evolution see \citet{Williams2011}.

The detection and characterisation of circumstellar discs was a primary goal of the 'Spitzer Space Telecope' \citep{Werner2004} and successfully featured in the Legacy Programme 'From Molecular Cores to Planet-Forming Dust' (c2d; \citet{Evans2009}). \citet{Allen2004} compared the predicted colours of young stellar objects with the photometric data obtained with the 'Infrared Array Camera' (IRAC; \citet{Fazio2004}). Transitional discs in different young clusters and associations were identified and classified by \citet{Muzerolle2010} by using IRAC data and the 24\,$\mu$m-fluxes obtained with MIPS \citep{Rieke2004}. \citet{Furlan2009}investigated the disc evolution in the three prominent star-forming regions Taurus, Ophiuchus, and Cha I and found similar transitional disc fractions in all three regions. This study is based on spectroscopic data taken with the 'Spitzer Infrared Spectrograph' \citep{Houck2004}. A spectroscopic study of the young stellar objects in Taurus has been published by \citet{Furlan2011}. The mid-infrared variability of (pre-)transitional disks was investigated with Spitzer (e.g.\,\citet{Espaillat2011}).

The interpretation of infrared colours in relation to the evolutionary state of a disc is, however, far from being trivial. This is particularly true with regards to the classification of transition discs, the latter being intended as objects caught in the act of dispersal, i.\,e.\,going from disc-bearing to disc-less stars. The modality and time-scales of disc dispersal are especially relevant to the planet formation process. In particular, inside-out dispersal from internal photoevaporation (e.g.\,\citet{Clarke2001, Alexander2006a, Alexander2006b, Owen2010, Owen2011b, Owen2011a, Owen2012a, Gorti2009}) could also provide a mechanism to stop migration and effectively 'park' planets at a particular location in a disc (e.g.\,\citet{Alexander2009}). 

The evolution of the dust component in a disc is mirrored by the evolution of colours in the infrared plane. By means of radiative transfer modelling, \citeauthor*{Ercolano2010} (2011, henceforth \citetalias{Ercolano2010}) identified the regions in the K\,-\,[8] vs. K\,-\,[24] plane where primordial discs, discs with inner-holes (i.\,e.\,presumably being dispersed from the inside-out) and discs which lose mass homogeneously at all radii, are expected to be found. Their study, which was limited to M-stars, showed that in the case of the cluster IC348, most discs disperse from the inside-out and undergo the transition on a short time-scale, as predicted by standard photoionisation models.

These conclusions are in contrast with the conclusions of \citet{Currie2009b}, who claimed instead a large number of 'homogeneously depleting' discs, for the same cluster. Such discrepancies highlight the need for detailed modelling in the interpretation of IR colours of discs.

The study of \citetalias{Ercolano2010} was restricted to M-stars in only one cluster, which prevented the authors from being able to make a more general statement with regards to disc dispersal. 

In this study we significantly improve on the work of \citetalias{Ercolano2010} by performing further RT calculation to evaluate evolutionary tracks in the IR colour plane for stars of different spectral types. We then apply our results to the photometric data of 15 nearby star-forming regions, that we collected from the literature, in order to address the question of what is the preferred mode of disc dispersal. 

In Section~\ref{Theoretical Investigation} we describe our methods and the results from our radiative transfer calculations and present new evolutionary diagnostic diagrams. In Section~\ref{Analysis and Discussion} the collected observational data are briefly described and analysed using the diagrams presented in the previous section. A short summary and conclusions are given in Section~\ref{Conclusion}. 

\section{Theoretical Investigation}
\label{Theoretical Investigation}

Emission in a given wavelength band is produced by dust grains of given temperatures and sizes. The temperature of grains in a disc for a given central star temperature is determined roughly by the grain location within the disc and by the intervening material. Changes in the distribution of material within a disc are therefore reflected in its emergent continuous spectral energy distribution (SED). On this basis, the evolution of infrared colours of an object can be used to trace the evolution of its dust density distribution. We extended the work of \citetalias{Ercolano2010} and calculate theoretical evolutionary tracks in the K\,-\,[8] vs. K\,-\,[24] plane by means of radiative transfer calculations for discs of different geometry and different evolutionary stage surrounding stars of different spectral types.

\subsection{Modes of Disc Dispersal}
\label{Modes of Disc Dispersal}

\begin{figure}
\centering
\includegraphics[width=0.3\textwidth]{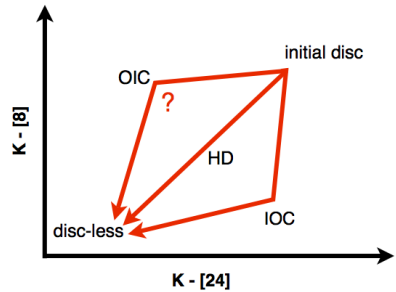} 
\caption{Expected tracks of discs depleting via the three different modes of disc dispersal explained in Section~\ref{Modes of Disc Dispersal}. The OIC track is marked with a question-mark, since it has been drawn according to the expectations from the simple thought experiment described in Section~\ref{Modes of Disc Dispersal}. But as will be discussed in Section~\ref{Razor thin discs}, the real tracks behave somewhat differently.}
\label{sketch}
\end{figure}

We consider three modes of disc dispersal: (i) homogeneous draining (HD), (ii) inside-out clearing (IOC) and (iii) outside-in clearing (OIC). The expected tracks in the infrared two-colour plane are sketched in Figure~\ref{sketch}.

\subsubsection{Homogeneous draining}
\label{HD}

HD refers to a disc whose surface density decreases homogeneously at all radii as a function of time, such that the surface density distribution $\Sigma(t)$ is self-similar at all times. Such a mode of dispersal would be expected if viscous draining were to act alone, undisturbed until all of the disc material is accreted onto the central object. In this case the total mass of the disc evolves roughly as $M_{\rm disc} \propto \Sigma(t) \propto t^{-1.5}$ \citep{Hartman1998}. In this scenario one would expect discs to evolve diagonally on the K\,-\,[8] vs. K\,-\,[24] plane (see Figure~\ref{sketch}) as both the 8\,$\mu$m and the 24\,$\mu$m band fluxes decrease roughly at the same rate as material is removed homogeneously at all radii. 

\subsubsection{Inside-out clearing}
\label{IOC}
IOC refers to discs that disperse via the formation of an inner-hole (classical transition discs), which becomes larger and larger until all material is removed. There is currently a healthy debate in the literature with regards to possible mechanisms that may be responsible for the formation of discs with inner-holes, of which a large variety is observed. Some of these mechanisms include photoevaporation by the central object (e.g.\,\citet{Alexander2006a, Alexander2006b, Owen2010, Owen2011b}), planet formation (e.g.\,\citet{Zhu2011}), grain growth (e.g.\,\citet{Dullemond2005}), close-binary effects (e.g.\,\citet{Ireland2008}) and external photoevaporation by nearby high-mass stars (e.g. Scally \& Clarke 2001). The picture that seems to be currently emerging is that of multiple populations of inner-hole objects, some of which may or may not represent objects in transition (e.g.\,\citet{Owen2012a,Owen2012b}). It is beyond the scope of this work to explore these different possibilities; here we will limit ourselves to calculating SEDs of discs, which, by whatever mechanism, develop increasingly larger inner-holes. The IOC tracks are expected to proceed in a left-facing 'L'-shape (see Figure~\ref{sketch}) as the warmer material in the inner disc is removed first, causing a drop in the 8\,$\mu$m flux, and the cooler material, responsible for most of the 24\,$\mu$m emission, is removed afterwards. 

\subsubsection{Outside-in clearing}
\label{OIC}
OIC refers to discs that are eroded from the outside-in by a companion star (e.g.\,\citet{Bouwman2006,Monin2007,Cieza2009}) or by radiation from external sources. External photo-evaporation is for example observed in the case of proplyds in the Orion Nebular Cluster (e.g.\,\citet{Ricci2008, McCaughrean1996,Scally2001}). Whether the entire mass of a disc can be dispersed via this mechanism remains, however, very uncertain. It is also uncertain what percentage of discs in OB associations and clusters are exposed to strong enough radiation from nearby massive stars for this process to be significant. In any case, it is an interesting exercise here to explore the evolutionary tracks for such a population on the infrared colour-colour plane. One would naively expect an upside-down 'L'-shape this time, almost a mirror-image of the IOC tracks (see Figure~\ref{sketch}), as the 24\,$\mu$m flux should drop before the 8\,$\mu$m flux. This is, however, not necessarily the case, as will be discussed in Section~\ref{Razor thin discs}.

\subsection{Numerical Set-up}
\label{Numerical Set-up}

We use the 2-dimensional Monte Carlo dust radiative transfer code of \citet{Whitney2003b,Whitney2003a}, which is routinely applied to the modelling of the SEDs of protoplanetary discs (e.\,g. \citet{Robitaille2006,Robitaille2007}, \citetalias{Ercolano2010}). We derive evolutionary tracks for stars of spectral types K4, K8.5, M4.5 and M8, using the stellar atmosphere models (\citet{Castelli2004, Brott2005}) from the library used by \cite{Robitaille2006} and available online at \verb|http://caravan.astro.wisc.edu/protostars/|. Table~\ref{spectral types} summarises the stellar parameters and provides indices for the atmosphere files used. Table~\ref{parameters} lists the other input parameters used in the 1287 radiative transfer models. A detailed description of the geometrical set-up and the input parameters is given in \cite{Robitaille2006}. 

 \begin{table}
 \caption{Spectral type intervals and stellar properties.}
 \label{spectral types}
  \begin{minipage}[b]{20cm} 
\begin{tabular}{clrSS}
 \hline
\multirow{2}{*}{index}&\multirow{2}{*}{SpT}&\multirow{2}{*}{$T_{\star}[K]$}&\multirow{2}{*}{$R_{\star} [R_{\sun}]$}&\multirow{2}{*}{$M_{\star} [M_{\sun}]$}\\
&&&&\\
\hline
3013427&K4&4590&1.35&1.21\\
3014686&K8.5&4060&3.73&0.75\\
3002746&M4.5&3305&0.844&0.228\\
3014449&M8&2805&1.8&0.1\\
\hline
 \end{tabular}
 \end{minipage}
 \end{table}

 \begin{table}
 \caption{Input parameter ranges used in our simulations\newline (see also \citet{Whitney2003b,Whitney2003a} and \citet{Robitaille2006}\newline for a detailed description).}
 \label{parameters}
  \begin{minipage}[b]{20cm} 
\begin{tabular}{lrr}
 \hline
\multirow{2}{*}{parameter}&\multirow{2}{*}{range}&\multirow{2}{*}{sequence}\\
&&\\
\hline
$M_{disc}$	&$10^{-2}\,M_{\star}$		&initial, IOC, OIC\\
$M_{disc}$	&$(10^{-2}-10^{-8})\,M_{\star}$&HD\\
$r_{disc}(max)$&$100\,AU$				&initial, IOC, HD\\
$r_{disc}(max)$&$(0.02-100)\,AU$			&OIC\\
$r_{disc}(min)$	&$1\,R_{sub}$\footnote{sublimation radius}				&initial, OIC, HD\\
$r_{disc}(min)$	&$1 - 30\,R_{sub}$			&IOC \footnote{razor-thin}\\
$r_{disc}(min)$	&$1 - 5000\,R_{sub}$			&IOC \footnote{finite thickness}\\
$H_{D}$\footnote{disc scale-height, see also Table~\ref{disc properties}}&0.0001 - 1.0&initial, IOC, OIC, HD\\
$A$\footnote{disc density exponent ($\sim r^{-\alpha}$)}&$2.0 - 2.25$&initial, IOC, OIC, HD\\
$B$\footnote{disc scale height exponent ($\sim r^{\beta}$); see also Table~\ref{disc properties}}&$1.0 - 1.25$&initial, IOC, OIC, HD\\
$\alpha_{disc}$\footnote{effectively no viscous heating}&$10^{-10}$&initial, IOC, OIC, HD\\
Mass infall rate\footnote{effectively no envelope}&$10^{-15}M_{\odot}/yr$&initial, IOC, OIC, HD\\
Aperture&100000\,AU&initial, IOC, OIC, HD\\
\hline
 \end{tabular}
 \end{minipage}
 \end{table}

We follow \citetalias{Ercolano2010} and consider several disc geometries exploring first the evolution of razor-thin discs for different dispersal modes and then extend our study to discs of finite thickness (Table~\ref{disc properties}). 

\subsection{Results}
\label{Results}

\subsubsection{Razor-thin discs}
\label{Razor thin discs}

\begin{figure*}
\centering
\includegraphics[width=\textwidth]{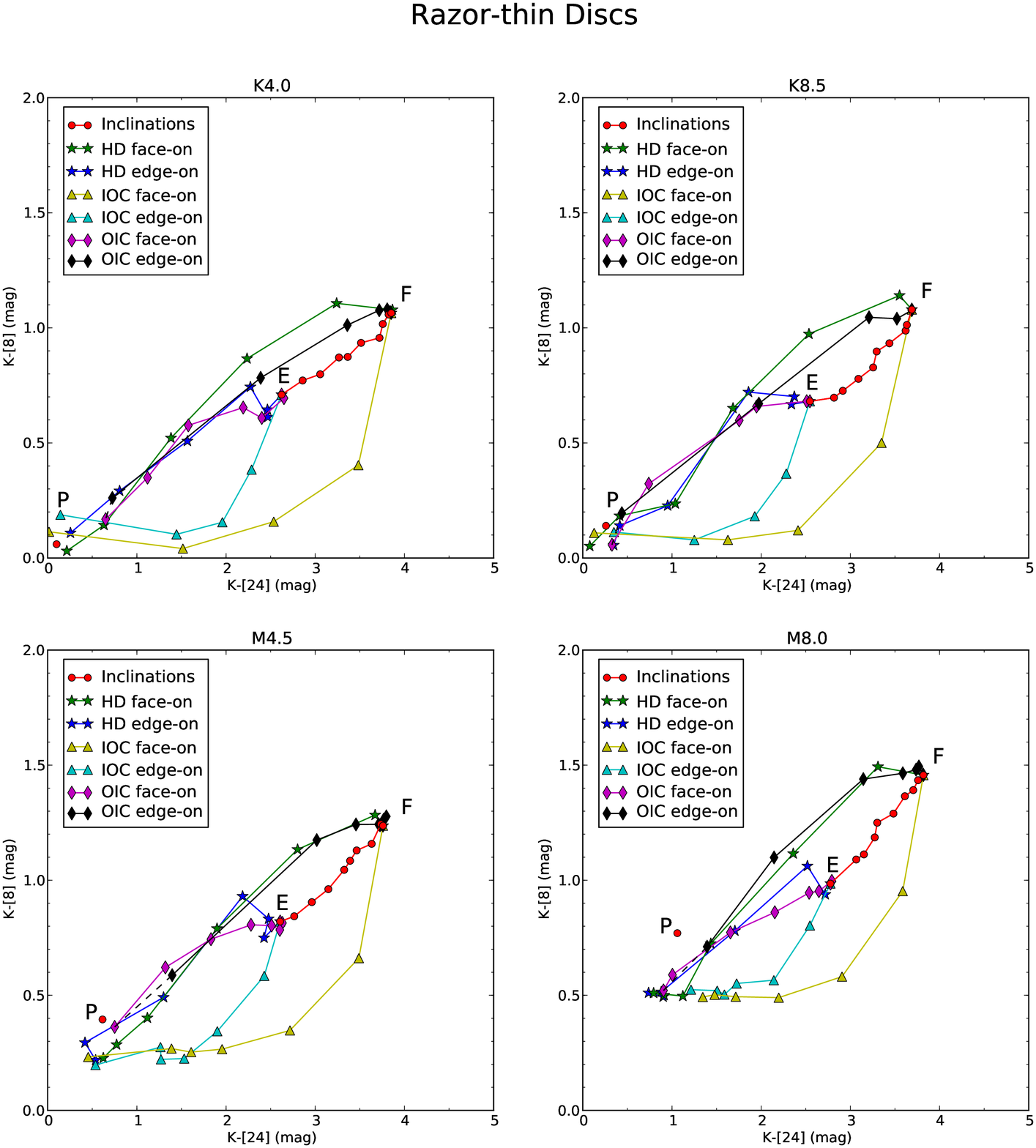} 
\caption{Evolution of a razor-thin disc via HD, IOC and OIC for all four spectral type intervals as described in Section~\ref{Modes of Disc Dispersal} and analysed in Section~\ref{Razor thin discs}. Point F are face-on ($\cos{i}=0.95$) and point E are edge-on ($\cos{i}=0.05$) disc inclinations. The points between the F and E represent intermediate disc inclinations in steps of $\Delta\cos{i}=0.10$. The red point P represents the approximate locus of the stellar photosphere as given by \citet{Luhman2010}. The stars mark the HD tracks for face-on and edge-on inclinations (green and blue lines, respectively). The triangles mark the IOC tracks for face-on and edge-on inclinations (yellow and cyan lines, respectively). The diamonds mark the OIC tracks for face-on and edge-on inclinations (black and purple lines, respectively).}
\label{thin disk OIC}
\end{figure*}

Figure~\ref{thin disk OIC} summarises the tracks for razor-thin discs in the four spectral type intervals considered here. Point F are face-on ($\cos{i}=0.95$) and point E are edge-on ($\cos{i}=0.05$) disc inclinations. The points between the F and E represent intermediate disc inclinations in steps of $\Delta\cos{i}=0.10$. Point P shows the approximate location of disc-less objects. The stars mark the HD tracks for face-on and edge-on inclinations (green and blue lines, respectively). The triangles mark the IOC tracks for face-on and edge-on inclinations (yellow and cyan lines, respectively). The diamonds mark the OIC tracks for face-on and edge-on inclinations (black and purple lines, respectively).

As expected (discussed in Section~\ref{HD} and sketched in Figure~\ref{sketch}), the HD tracks follow a diagonal trajectory from points F and E to P, while the IOC tracks show the typical left-facing 'L'-shape also reported by \citetalias{Ercolano2010} for M-star models. The OIC tracks in Figure~\ref{thin disk OIC}, however, do not show the upside-down 'L'-shape suggested by our simple-minded thought experiment (see Section~\ref{OIC}). This is because the 24\,$\mu$m emission is not dominated by material at large radii in the disc, but contains a significant contribution from material in the inner disc that is located deeper inside and is hence cooler. 

\subsubsection{Discs with finite-thickness}
\label{Discs with finite thickness}

 \begin{table}
 \caption{Modelling parameters of primordial discs.}
 \label{disc properties}
  \begin{minipage}[b]{15cm} 
\begin{tabular}{cSrrrSrrr}
&\multicolumn{4}{c}{flaring}&\multicolumn{4}{c}{mixing}\\
\hline
\multirow{2}{*}{disc kind}	&\multirow{2}{*}{\   B}&\multirow{2}{*}{\footnote{constant opening angle}}&\multirow{2}{*}{\footnote{flared}}&\multirow{2}{*}{\footnote{strongly flared}}&\multirow{2}{*}{ \ \ \  $H_{D}$}	&\multirow{2}{*}{\footnote{strongly settled}}&\multirow{2}{*}{\footnote{settled}}&\multirow{2}{*}{\footnote{mixed}}\\
&&\multicolumn{3}{c}{$\longrightarrow$}&&\multicolumn{3}{c}{$\longrightarrow$}\\
\hline
razor-thin&1.&x&&&0.0001&x&&\\
\hline
ultra-settled&1.		&x&&&0.1	&x&&\\
\multirow{2}{*}{optically thick}&1.		&x&&&0.5	&&x&\\
&1.		&x&&&1.	&&&x\\
ultra-settled&1.13	&&x&&0.1	&x&&\\
\multirow{2}{*}{optically thick}&1.13	&&x&&0.5	&&x&\\
&1.13	&&x&&1.	&&&x\\
ultra-settled&1.25	&&&x&0.1	&x&&\\
\multirow{2}{*}{optically thick} &1.25	&&&x&0.5	&&x&\\
&1.25	&&&x&1.	&&&x\\
\hline
 \end{tabular}
 \end{minipage}
 \end{table}

\begin{figure*}
\centering
\includegraphics[width=\textwidth]{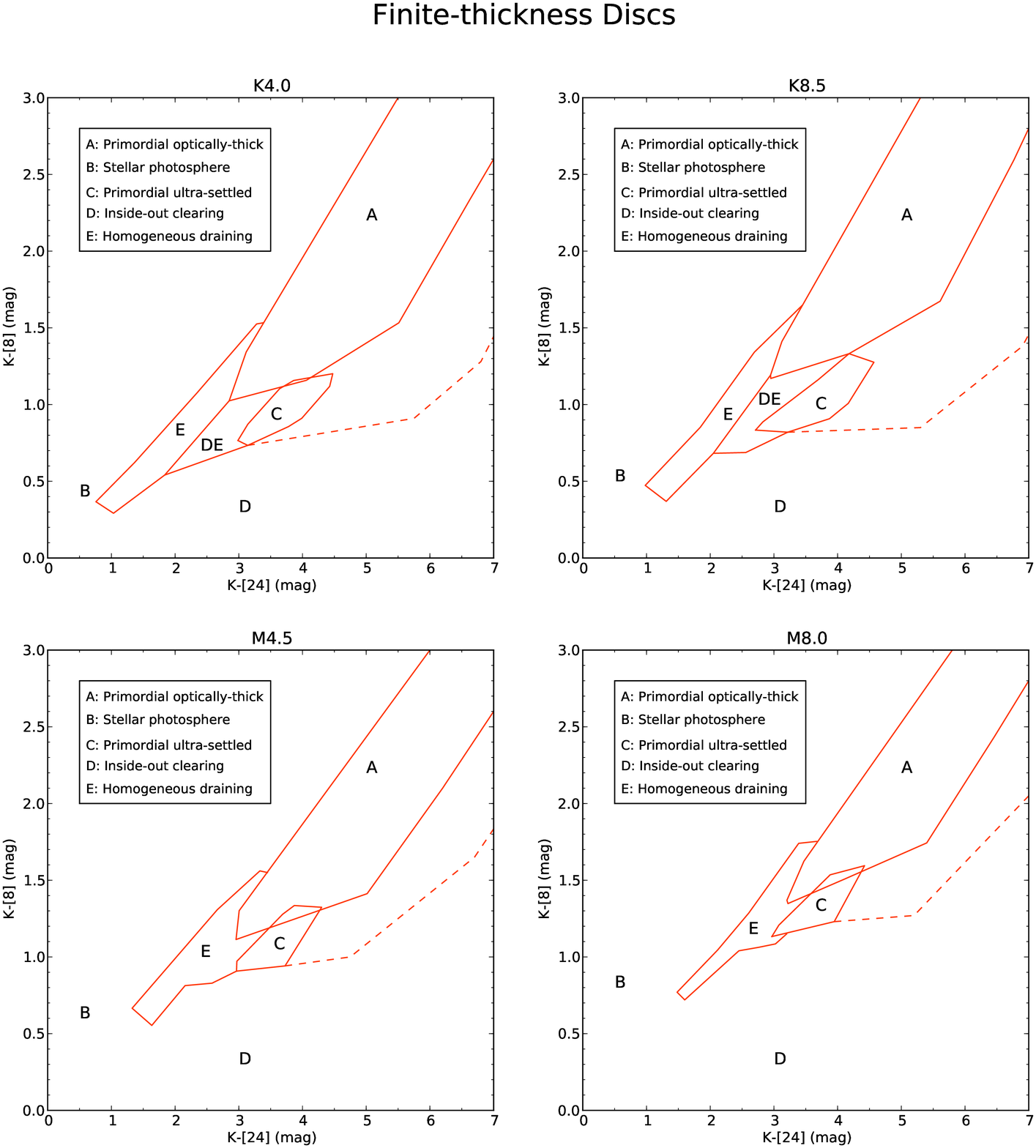} 
\caption{K\,-\,[8] vs. K\,-\,[24] evolutionary diagnostic diagrams (dashed lines represent objects with envelopes). \newline The values of the boundaries are given in the Appendix.}
\label{thick disk}
\end{figure*}

We follow \citetalias{Ercolano2010} and set-up discs of finite-thickness and nine different geometries; we extend, however, the work of \citetalias{Ercolano2010} to all the spectral types considered here. The resulting infrared colours are plotted in Figure~\ref{thick disk} for all cases where disc extinction $A_V < 10$\, mag. The parameter set-up is summarised in Table~\ref{disc properties}.

We explore three different cases of dust settling (by varying the parameter $H_D$) and three different levels of flaring, in order to cover the parameter space occupied by primordial discs of different geometries. Discs with $H_D = 0.1$ are named 'primordial ultra-settled', while those with a larger $H_D\geq0.5$ are named 'primordial optically-thick' discs. We note that $H_D$ also influences the inner wall height, which may be the main source of variation of the NIR/~MIR colours.

By inspection of the locus occupied by the evolutionary tracks of primordial (ultra-settled and optically-thick) discs, we define boundary boxes in the two-colour plane. Our results are summarised in Figure~\ref{thick disk}. As for the razor-thin case, we carried out radiative transfer calculations to track the colour evolution of each of our primordial (ultra-settled or optically-thick) discs as they disperse via HD and IOC, allowing us to further identify the expected locus of objects in such evolutionary stages. For the sake of clarity, the tracks are not plotted in Figure~\ref{thick disk}, where only the derived boundary boxes are shown. In particular the boxes are labelled as A, B, C, D, E for 'primordial optically-thick' discs, disc-less sources, 'primordial ultra-settled' discs, IOC and HD respectively. Due to the overlap in the evolving discs regimes D and E, we cannot discriminate between HD and IOC for those objects observed in DE. As will be shown later, however, only a very small minority of the observational sample lies in this regime. 

	Additionally, we run a set of models to allow us to constrain the locus of envelope objects and estimate the contamination of the IOC objects by this sample. We set up 576 objects with envelopes (see \citet{Whitney2003b}) for a selection of cavity angles ($10^\circ$ to $90^\circ$) and envelope infall rates ($10^{-6}M_{\odot}/yr$ to $10^{-9}M_{\odot}/yr$). Objects with envelopes and high infall rates show the greatest overlap with the IOC region, whereas objects with less infall are bluer. We have indicated the locus of these models with a dashed line in Figure~\ref{thick disk} above which envelope objects lie. By comparing with observational data (see Section~\ref{Analysis and Discussion}), we found, that primordial discs with envelopes might introduce a contamination in the IOC statistics of at most 10\,\%. This bias has to be kept in mind when considering the raw statistics of the upcoming section. The boundary points of each box are listed in Appendix~\ref{appendix: boxes}. 

\section{Analysis and Discussion}
\label{Analysis and Discussion}
Our classification scheme, based on the radiative transfer modelling described in Section~\ref{Numerical Set-up}, allows us to use currently available infrared photometric surveys from nearby star-forming regions in order to address the question of what is the preferential mode of disc dispersal. We have collected K-band, 8\,$\mu$m and 24\,$\mu$m band data, as well as spectral types and extinction measurements of YSOs in 15 nearby star-forming regions. Table~\ref{observations} lists the references for the data and summarises some of the physical properties of the regions. 

\begin{table*}
 \caption{Overview of the used observational data.}
 \label{observations}
 \begin{minipage}[b]{15cm} 
 \renewcommand{\thempfootnote}{\arabic{mpfootnote}}
  \begin{tabular}{lcrrrrrrr}
  \hline
\multirow{2}{*}{region} & \multirow{2}{*}{$N_{tot}$} & \multirow{2}{*}{age [Myr]} & \multirow{2}{*}{distance [pc]}  & \multirow{2}{*}{JHK} & \multirow{2}{*}{[8]} & \multirow{2}{*}{[24]} & \multirow{2}{*}{$A_{\lambda}$} & \multirow{2}{*}{SpT}\\
&&&&&&&&\\
\hline
Taurus$^{L}$&211&1\footnotemark[1]$^{,}$\footnotemark[2]$^{,}$\footnotemark[3]&140\footnotemark[4]&\footnotemark[5]&\footnotemark[2]&\footnotemark[2]&\footnotemark[6]&\footnotemark[2]$^{,}$\footnotemark[6]\\
\hline
Taurus$^{R}$&257&1\footnotemark[1]$^{,}$\footnotemark[2]$^{,}$\footnotemark[3]&137\footnotemark[7]&\footnotemark[5]&\footnotemark[8]&\footnotemark[8]&\footnotemark[8]&\footnotemark[8]\\
\hline
NGC2068/71&27&2\footnotemark[9]&400\footnotemark[9]&\footnotemark[9]&\footnotemark[9]&\footnotemark[9]&\footnotemark[9]&\footnotemark[9]\\
\hline
NGC2264&313&2\footnotemark[10]&760-913\footnotemark[11]$^{,}$\footnotemark[12]&\footnotemark[13]&\footnotemark[13]&\footnotemark[13]&\footnotemark[13]&\footnotemark[13]\\
\hline
ChaI&29&2-3\footnotemark[2]$^{,}$\footnotemark[14]$^{,}$\footnotemark[15]&162\footnotemark[15]&\footnotemark[16]&\footnotemark[16]&\footnotemark[16]&\footnotemark[17]&\footnotemark[17]\\
\hline
IC348&240&2-3\footnotemark[18]$^{,}$\footnotemark[19]$^{,}$\footnotemark[20]&216-340\footnotemark[18]$^{,}$\footnotemark[20]$^{,}$\footnotemark[21]$^{,}$\footnotemark[22]$^{,}$\footnotemark[23]$^{,}$\footnotemark[24]&\footnotemark[21]&\footnotemark[21]&\footnotemark[21]&\footnotemark[21]&\footnotemark[21]\\
\hline
NGC1333&39&$<$3\footnotemark[25]&260\footnotemark[26]&\footnotemark[16]$^{,}$\footnotemark[25]&\footnotemark[16]$^{,}$\footnotemark[25]&\footnotemark[16]$^{,}$\footnotemark[25]&-&\footnotemark[25]\\
\hline
Tr37&57&4\footnotemark[27]$^{,}$\footnotemark[28]$^{,}$\footnotemark[29]&900\footnotemark[30]&\footnotemark[28]&\footnotemark[29]&\footnotemark[29]&\footnotemark[28]&\footnotemark[28]\\
\hline
Serpens&69&2-6\footnotemark[31]&260-415\footnotemark[32]$^{,}$\footnotemark[33]&\footnotemark[34]&\footnotemark[34]&\footnotemark[34]&\footnotemark[31]&\footnotemark[31]\\
\hline 
LupIII&82&2-6\footnotemark[35]$^{,}$\footnotemark[36]&200\footnotemark[35]&\footnotemark[34]&\footnotemark[34]&\footnotemark[34]&\footnotemark[36]&\footnotemark[36]\\
\hline
ChaII&27&4-5\footnotemark[37]&178\footnotemark[15]&\footnotemark[34]&\footnotemark[34]&\footnotemark[34]&\footnotemark[37]&\footnotemark[37]\\
\hline
NGC2362&10&5\footnotemark[38]&1480\footnotemark[14]&\footnotemark[14]&\footnotemark[14]&\footnotemark[14]&\footnotemark[39]&\footnotemark[39]\\
\hline
OB1bf&13&5\footnotemark[40]&440\footnotemark[41]&\footnotemark[5]&\footnotemark[42]&\footnotemark[42]&-&\footnotemark[43]\\
\hline
UpperSco&133&5\footnotemark[44]&145\footnotemark[45]&\footnotemark[5]&\footnotemark[46]&\footnotemark[47]&\footnotemark[47]&\footnotemark[47]\\
\hline
$\eta$ Cha&14&6\footnotemark[48]&100\footnotemark[48]&\footnotemark[5]&\footnotemark[2]&\footnotemark[2]&\footnotemark[1]&\footnotemark[49]\\
\hline
25Ori&8&7-10\footnotemark[50]&330\footnotemark[51]&\footnotemark[5]&\footnotemark[42]&\footnotemark[42]&-&\footnotemark[43]\\
\hline
 \end{tabular}
\begin{minipage}[b]{20cm}
L: Data from \cite{Luhman2010}\\
R: Data from \cite{Rebull2010}
\end{minipage}
\hfill
\begin{minipage}[b]{20cm}
   \begin{minipage}[b]{5cm}
\footnotetext[1]{\citet{Luhman2004b}}
\footnotetext[2]{\citet{Luhman2010}}
\footnotetext[3]{\citet{North}}
\footnotetext[4]{\citet{Kenyon1994}}
\footnotetext[5]{\citet{Cutri2003}}
\footnotetext[6]{\citet{Luhman2009}}
\footnotetext[7]{\citet{Torres2007}}
\footnotetext[8]{\citet{Rebull2010}}
\footnotetext[9]{\citet{Flaherty2008}} 
\footnotetext[10]{\citet{Teixeira2008}} 
\footnotetext[11]{\citet{Sung1997}}
\footnotetext[12]{\citet{Baxter2009}}   
\footnotetext[13]{\citet{Teixeira2012}} 
\footnotetext[14]{\citet{Currie2009a}}
\footnotetext[15]{\citet{Whittet1997}}
\footnotetext[16]{\citet{Gutermuth2009}}
\footnotetext[17]{\citet{Luhman2007}}
\end{minipage}
   \begin{minipage}[b]{5cm}
\footnotetext[18]{\citet{Herbig1998}}
\footnotetext[19]{\citet{Muench2003}}
\footnotetext[20]{\citet{Luhman2003}}
\footnotetext[21]{\citet{Currie2009b}}
\footnotetext[22]{\citet{Herbst2008}}
\footnotetext[23]{\citet{Scholz1999}}
\footnotetext[24]{\citet{Cernis1993}}
\footnotetext[25]{\citet{Winston2009}}
\footnotetext[26]{\citet{Hirota2008}}
\footnotetext[27]{\citet{Sicilia2004}}  
\footnotetext[28]{\citet{Sicilia2005}} 
\footnotetext[29]{\citet{Sicilia2006}}
\footnotetext[30]{\citet{Contreras2002}}
\footnotetext[31]{\citet{Oliveira2009}}
\footnotetext[32]{\citet{Straizys1996}}
\footnotetext[33]{\citet{Dzib2010}}
\footnotetext[34]{\citet{Evans2009}}
\end{minipage}
   \begin{minipage}[b]{5cm}
\footnotetext[35]{\citet{South}}
\footnotetext[36]{\citet{Merin2008}}
\footnotetext[37]{\citet{Spezzi2008}}
\footnotetext[38]{\citet{Mayne2008}}
\footnotetext[39]{\citet{Dahm2007}}
\footnotetext[40]{\citet{Hernandez2006}}
\footnotetext[41]{\citet{Hernandez2005}}
\footnotetext[42]{\citet{Hernandez2007a}}
\footnotetext[43]{\citet{Hernandezpersonal}}
\footnotetext[44]{\citet{Hernandez2007b}}
\footnotetext[45]{\citet{deZeeuw1999}}
\footnotetext[46]{\citet{Carpenter2006}}
\footnotetext[47]{\citet{Carpenter2009}}
\footnotetext[48]{\citet{Mamajek1999}}
\footnotetext[49]{\citet{LuhmanSteeghs2004}}
\footnotetext[50]{\citet{Briceno2007}}
\footnotetext[51]{\citet{Briceno2005}}
\end{minipage}
  \end{minipage}
  \end{minipage}
 \end{table*}

We applied our colour-colour diagnostic diagrams to classify 1529 objects in 15 nearby star-forming regions. The number statistics in each evolutionary stage and for each spectral type regime are presented in Table~\ref{statistics}. It is worth noticing here, that we deal in most cases with small number statistics; for this reason we provide actual numbers in Table~\ref{statistics} rather than percentages. In what follows, however, we prefer to use percentages in order to simplify the discussion. The standard statistical errors (e.g.\,$\sqrt{N}$), however, should be kept in mind when considering the raw percentages given. Given the overlap between regions D (IOC) and E (HD), discussed in the previous section, we can only provide upper and lower limits to the count sources in these regimes. 

In summary, 38.9\,\% of the objects out of the entire sample lie in the primordial disc region whereas between 31.3\,\%  and 31.9\,\% disperse their discs from the inside-out and up to 21.7\,\% of the sources have already lost their disc. So, almost a third of the available sources are currently clearing their discs from the inside-out. Only between 1.5\,\% and 2.0\,\%  of the objects lie in the homogeneous draining region E. We interpret this result as strong evidence against homogeneous disc depletion as the main disc dispersal mode.  

Figures~\ref{Taurus-Luhman} to~\ref{25Ori} show the results for individual star-forming regions, which are also briefly summarised in the next section. The regions are sorted by age.

\subsection{Analysis}
\label{Analysis}

\begin{figure*}
\centering
\includegraphics[width=\textwidth]{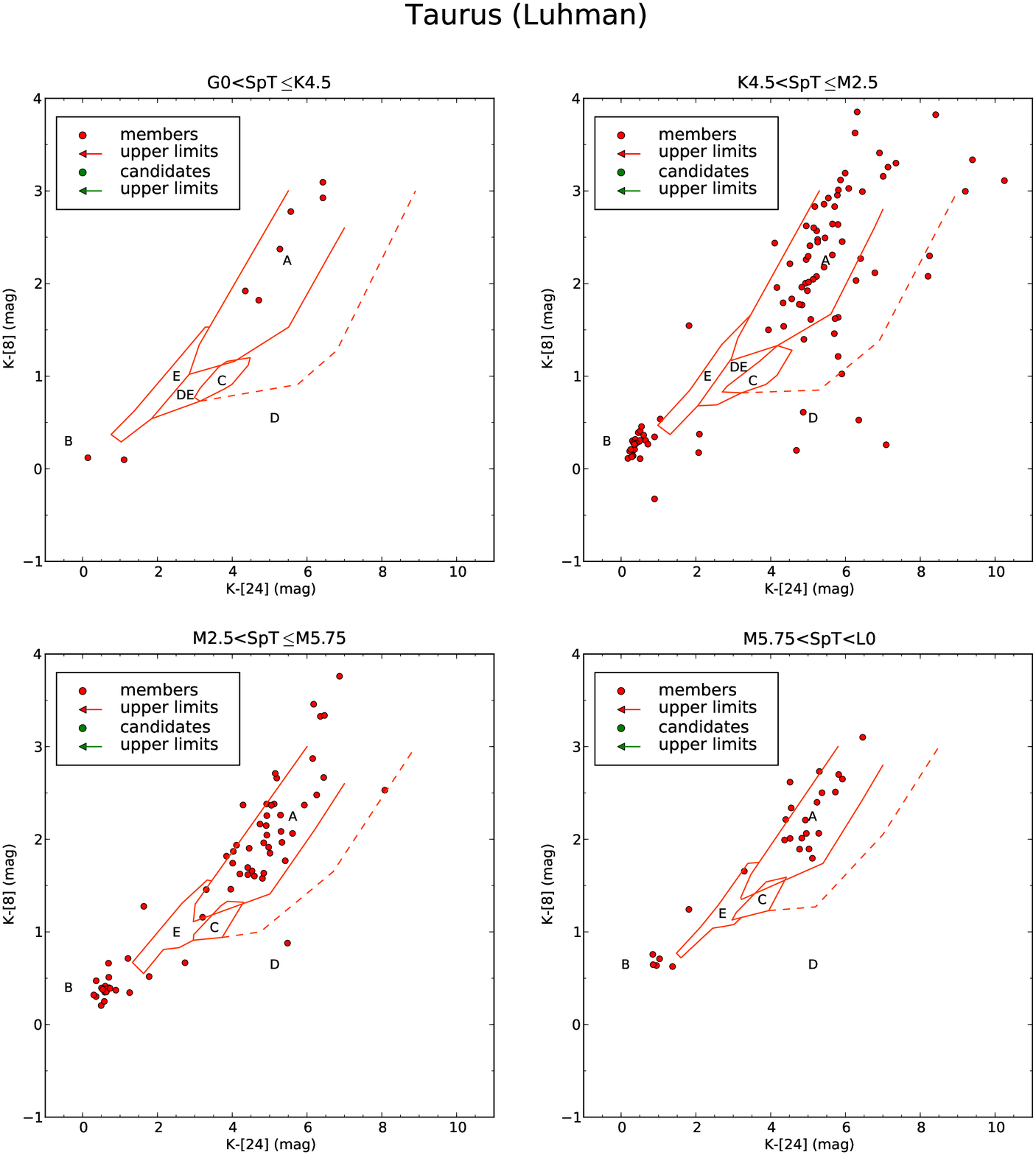} 
\caption{Disc evolution diagnostic diagram (dashed lines represent objects with envelopes) applied to the YSOs in the 1 Myr old cluster Taurus. 
(A:~51.2\,\% primordial optically-thick, C:~0.0\,\% primordial ultra-settled, B:~23.2\,\% disc-less, D:~13.7\,\% inside-out clearing, E:~0.5\,\% homogeneous draining) 20 sample points lie outside the limits of this plot, but they are still included in the final statistics.}
\label{Taurus-Luhman}
\end{figure*}

\begin{figure*}
\centering
\includegraphics[width=\textwidth]{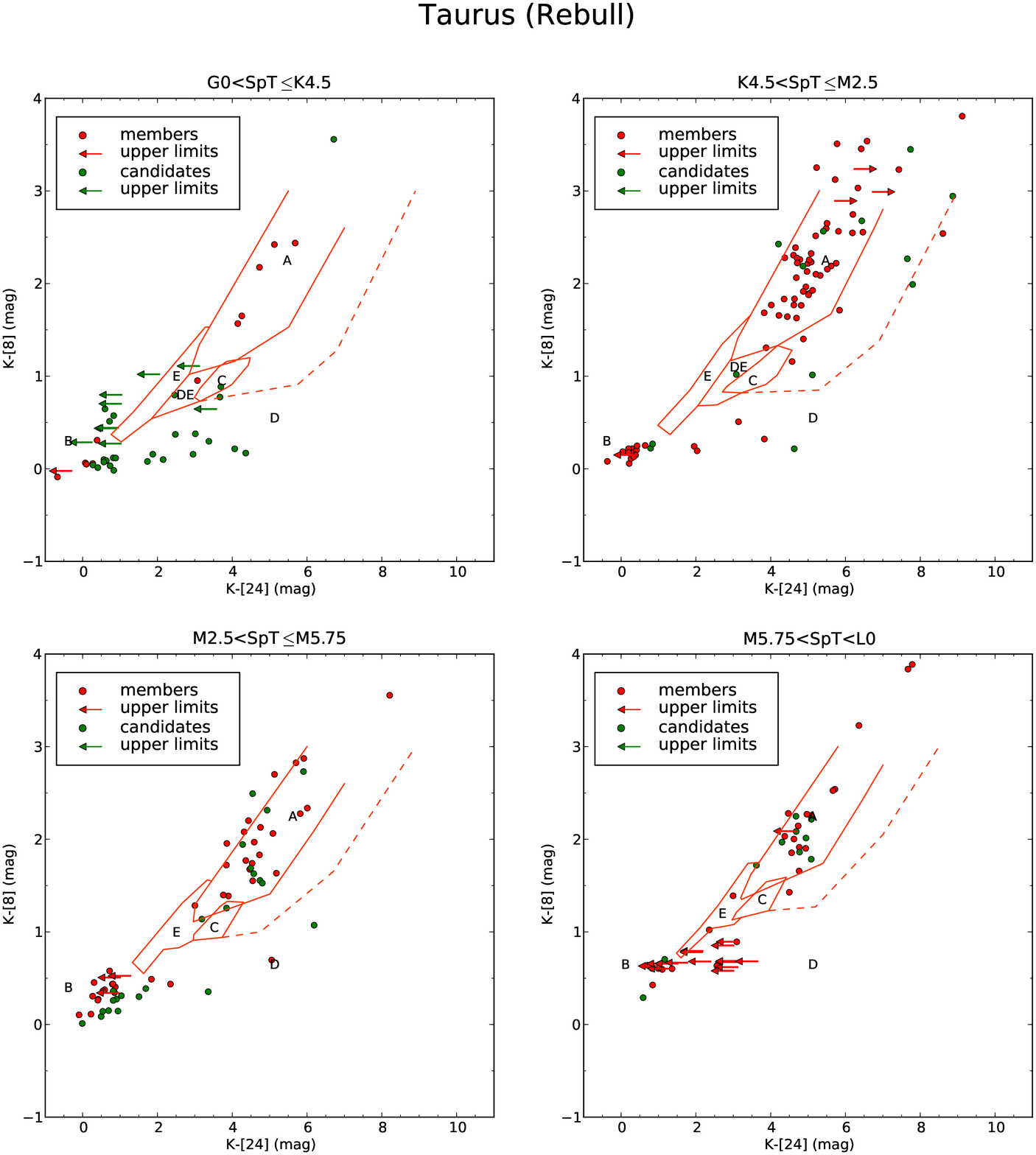} 
\caption{Disc evolution diagnostic diagram (dashed lines represent objects with envelopes) applied to the YSOs in the 1 Myr old cluster Taurus. 
(A:~41.6\,\% primordial optically-thick, C:~0.8\,\% primordial ultra-settled, B:~31.5\,\% disc-less, D:~(17.9) 19.1\,\% inside-out clearing, E:~(2.3) 3.5\,\% homogeneous draining) 5 sample points lie outside the limits of this plot, but they are still included in the final statistics.}
\label{Taurus-Rebull}
\end{figure*}

\begin{figure*}
\centering
\includegraphics[width=\textwidth]{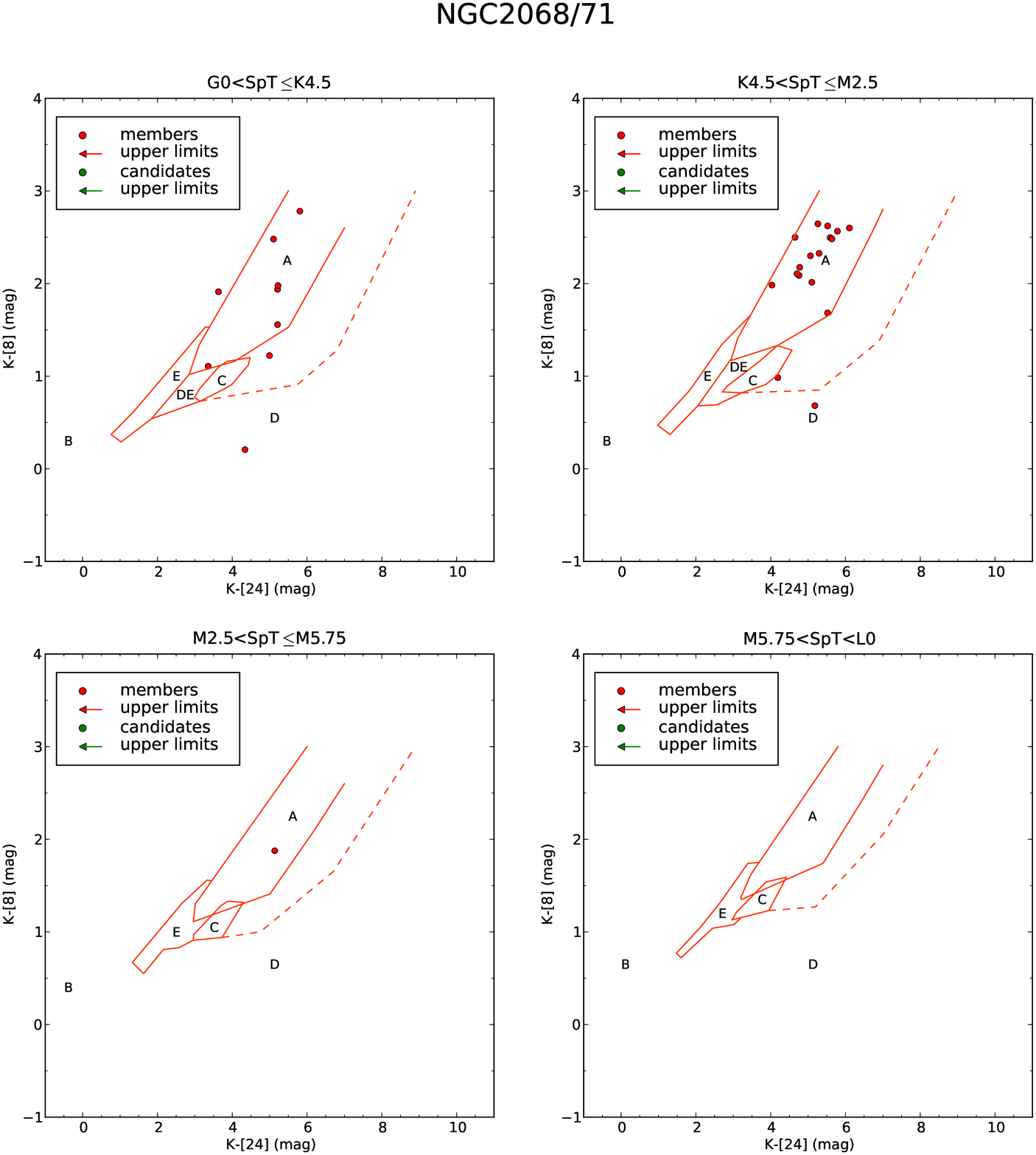} 
\caption{Disc evolution diagnostic diagram (dashed lines represent objects with envelopes) applied to the YSOs in the 2 Myr old cluster NGC2068/71. 
(A:~81.5\,\% primordial optically-thick, C:~0.0\,\% primordial ultra-settled, B:~0.0\,\% disc-less, D:~14.8\,\% inside-out clearing, E:~0.0\,\% homogeneous draining)}
\label{NGC2068-71}
\end{figure*}

\begin{figure*}
\centering
\includegraphics[width=\textwidth]{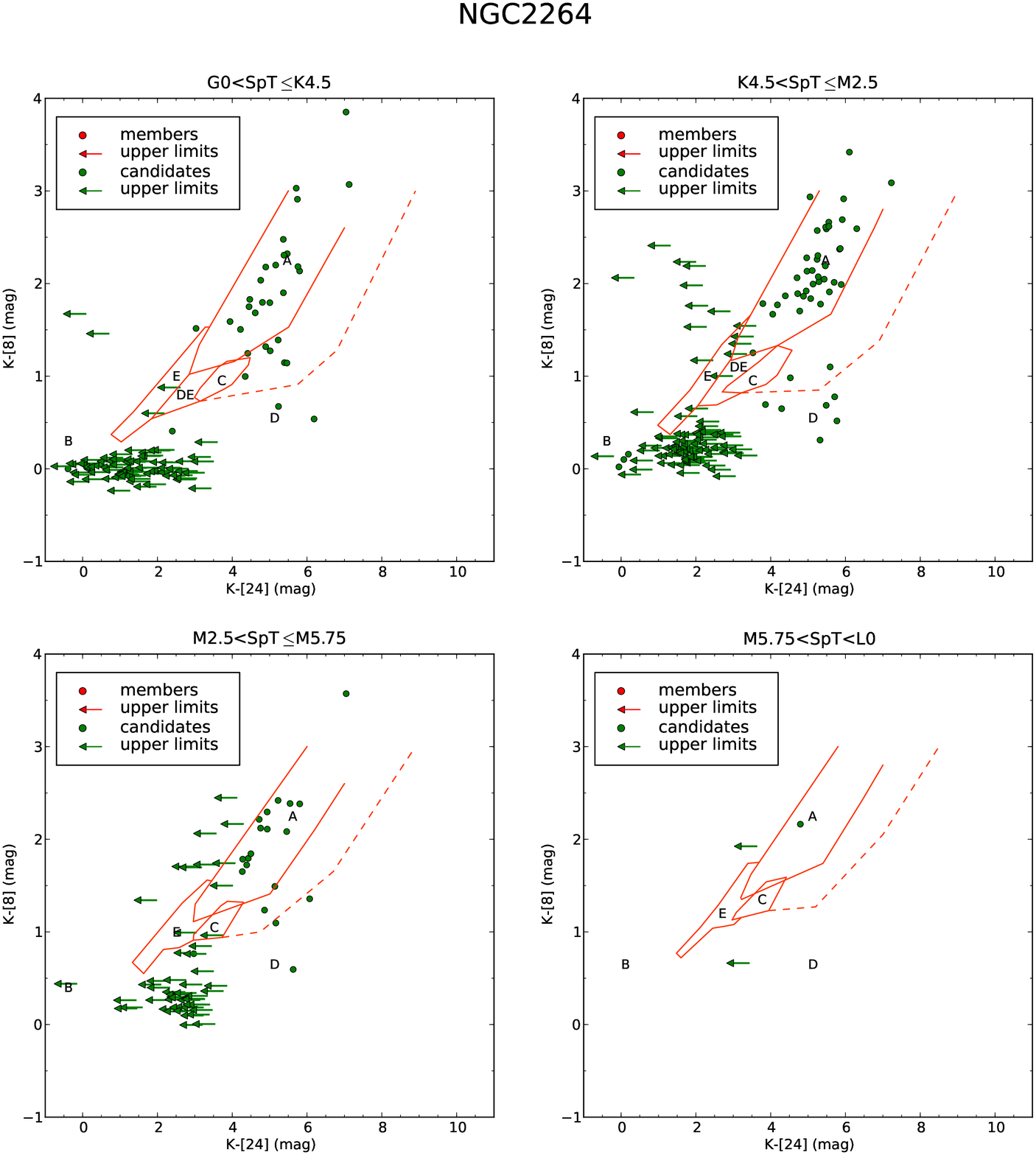} 
\caption{Disc evolution diagnostic diagram (dashed lines represent objects with envelopes) applied to the YSOs in the 2 Myr old cluster NGC2264. 
(A:~25.2\,\% primordial optically-thick, C:~0.6\,\% primordial ultra-settled, B:~9.9\,\% disc-less, D:~(56.2) 57.2\,\% inside-out clearing, E:~(0.3) 1.3\,\% homogeneous draining)}
\label{NGC2264}
\end{figure*}

\begin{figure*}
\centering
\includegraphics[width=\textwidth]{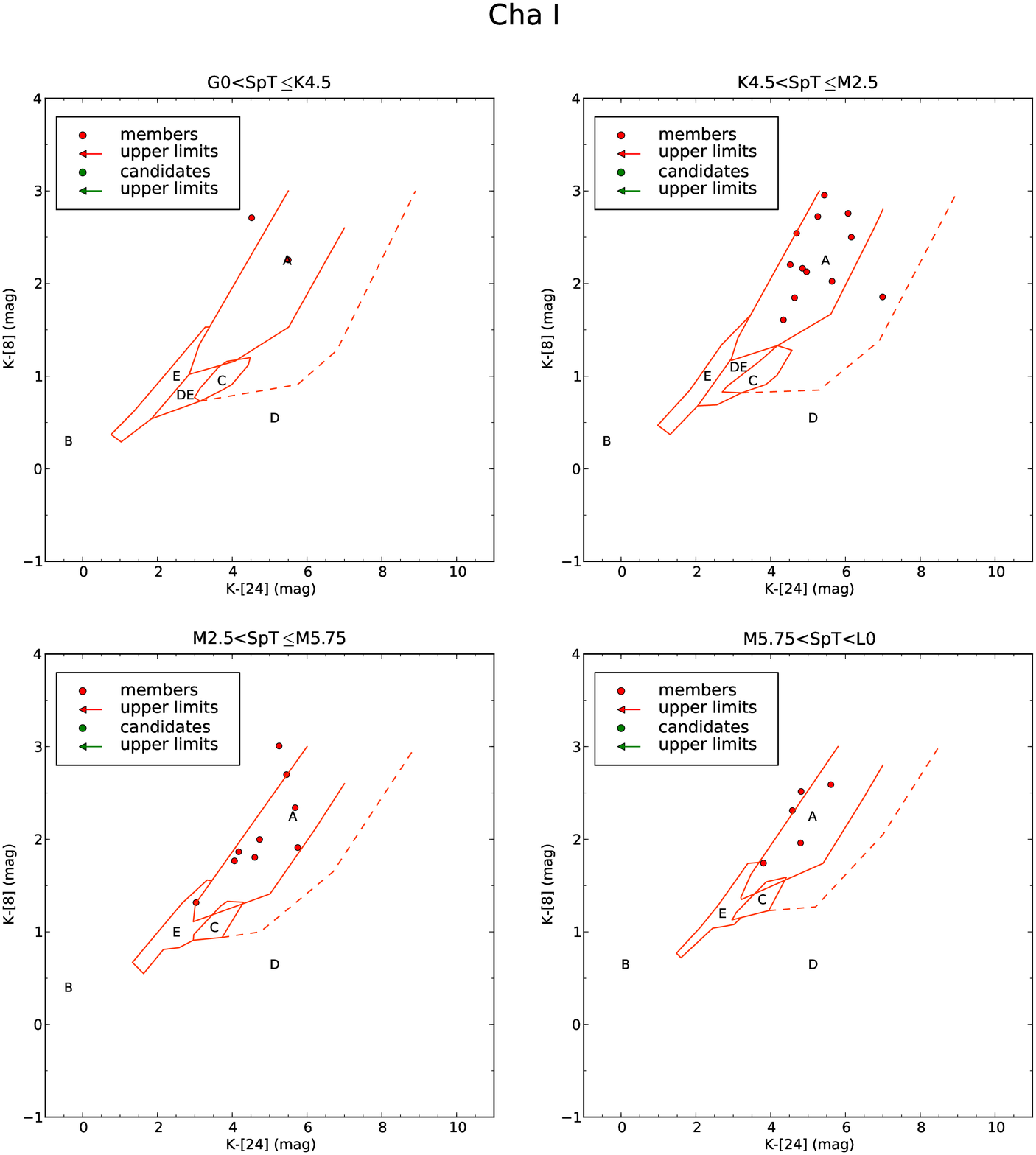} 
\caption{Disc evolution diagnostic diagram (dashed lines represent objects with envelopes) applied to the YSOs in the 2-3 Myr old cluster Cha I. 
(A:~75.9\,\% primordial optically-thick, C:~0.0\,\% primordial ultra-settled, B:~0.0\,\% disc-less, D:~3.4\,\% inside-out clearing, E:~3.4\,\% homogeneous draining) 1 sample point lies outside the limits of this plot, but it is still included in the final statistics.}
\label{ChaI}
\end{figure*}

\begin{figure*}
\centering
\includegraphics[width=\textwidth]{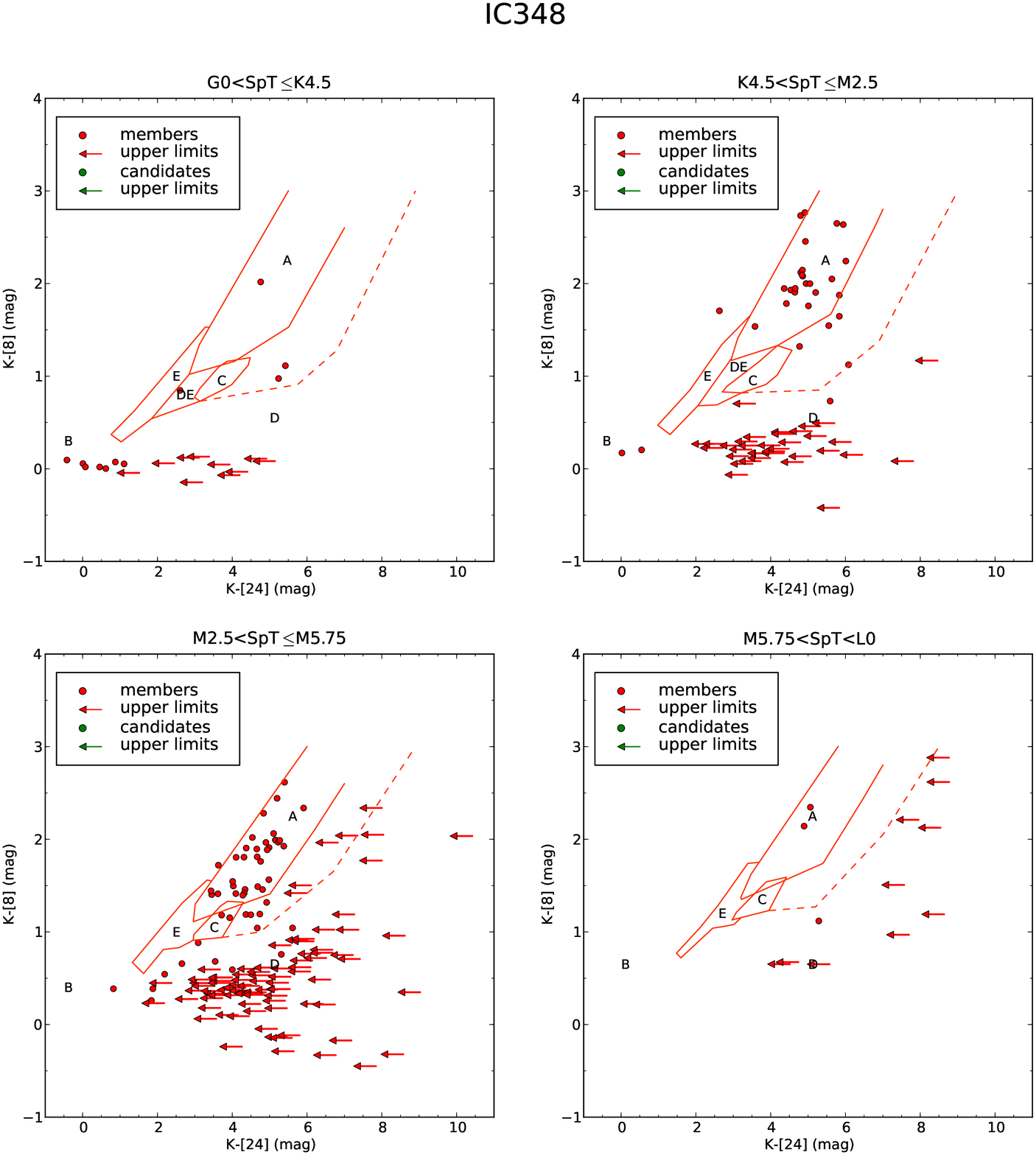} 
\caption{Disc evolution diagnostic diagram (dashed lines represent objects with envelopes) applied to the YSOs in the 2-3 Myr old cluster IC348. 
(A:~23.8\,\% primordial optically-thick, C:~0.8\,\% primordial ultra-settled, B:~3.8\,\% disc-less, D:~(69.6) 70.0\,\% inside-out clearing, E:~(0.0) 0.4\% homogeneous draining) 3 sample points lie outside the limits of this plot, but they are still included in the final statistics.}
\label{IC348}
\end{figure*}

\begin{figure*}
\centering
\includegraphics[width=\textwidth]{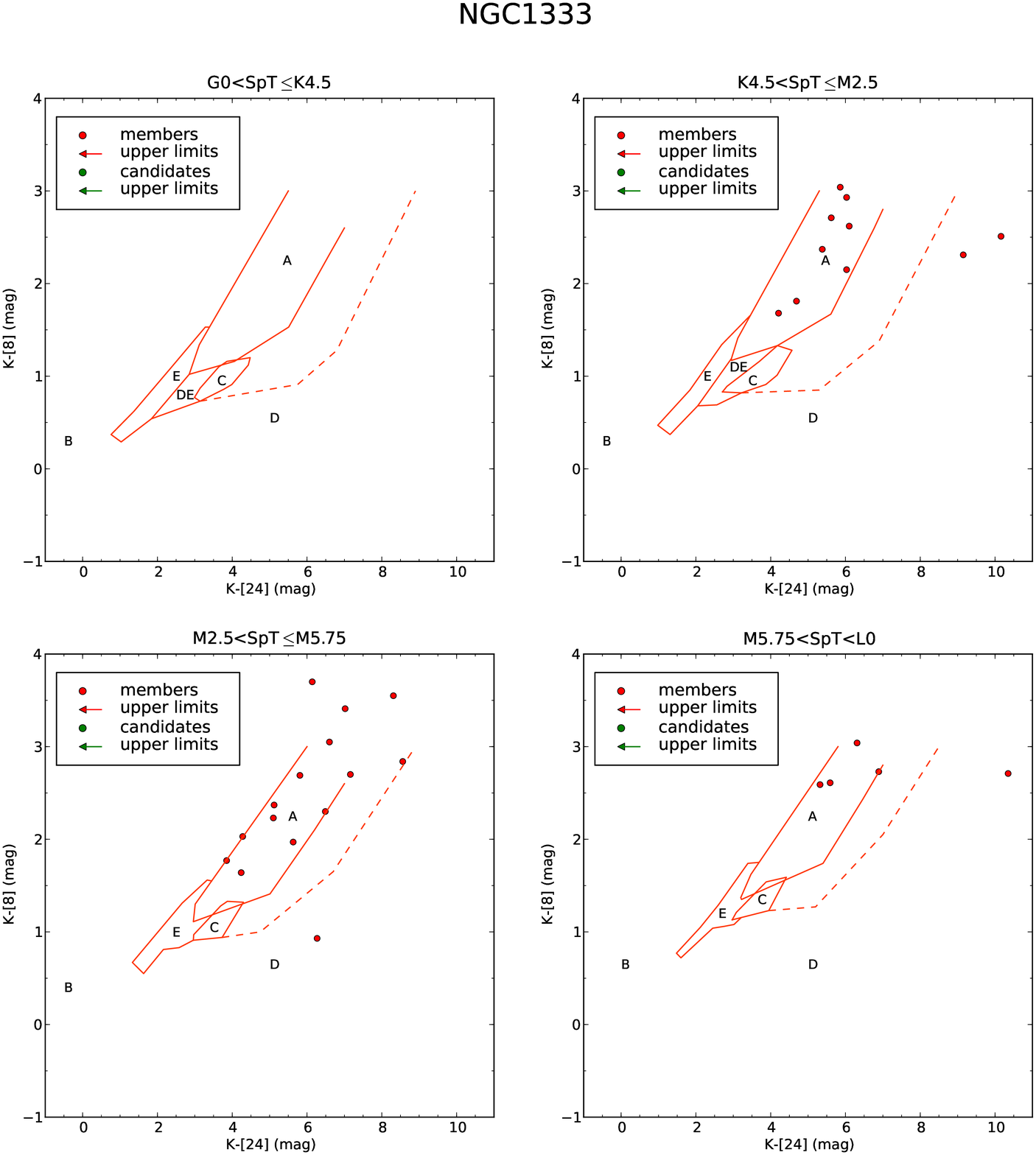} 
\caption{Disc evolution diagnostic diagram (dashed lines represent objects with envelopes) applied to the YSOs in the less than 3 Myr old cluster NGC1333. 
(A:~66.7\,\% primordial optically-thick, C:~0.0\,\% primordial ultra-settled, B:~0.0\,\% disc-less, D:~28.2\,\% inside-out clearing, E:~0.0\,\% homogeneous draining) 9 sample points lie outside the limits of this plot, but they are still included in the final statistics.}
\label{NGC1333}
\end{figure*}

\begin{figure*}
\centering
\includegraphics[width=\textwidth]{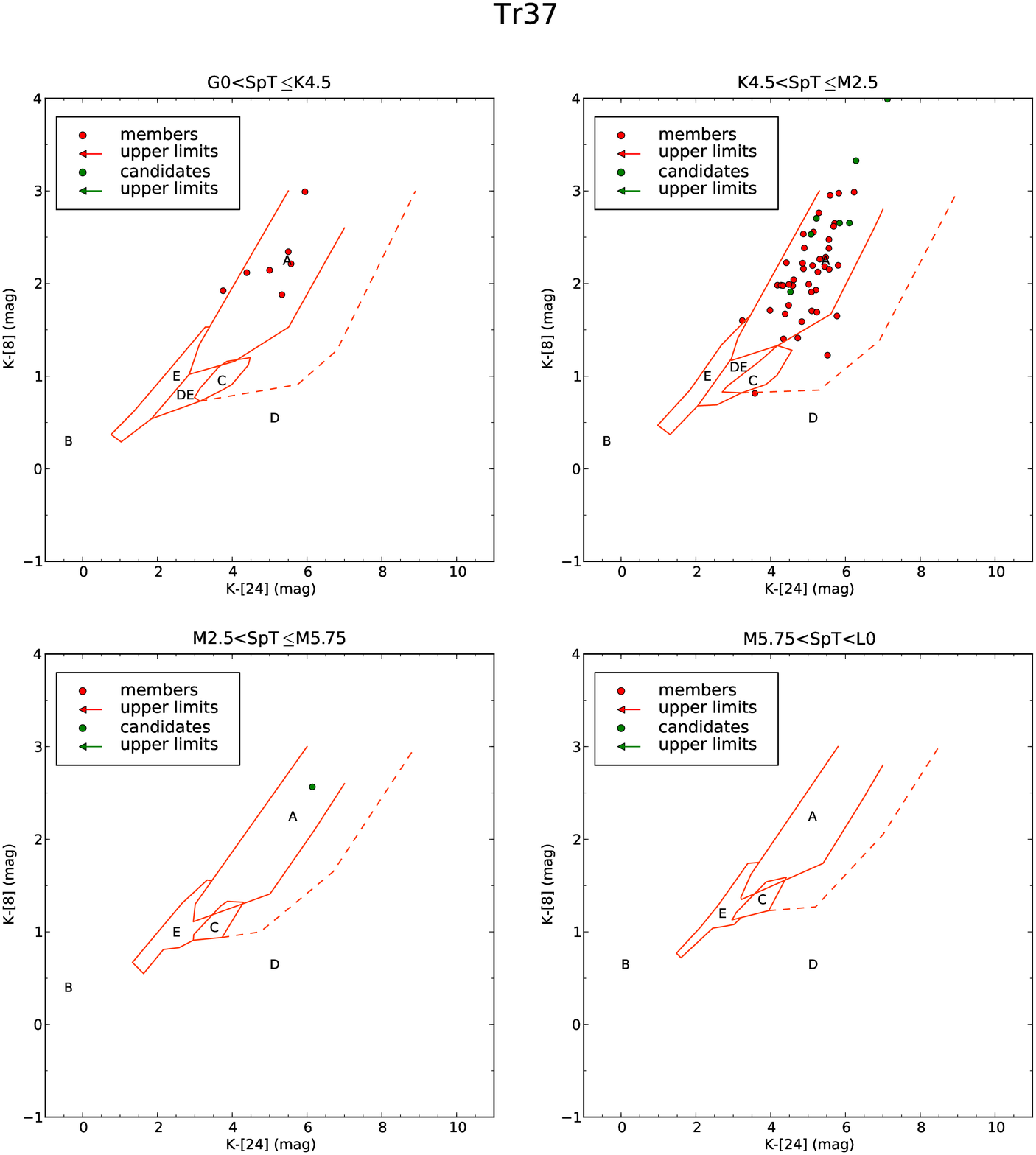} 
\caption{Disc evolution diagnostic diagram (dashed lines represent objects with envelopes) applied to the YSOs in the 4 Myr old cluster Tr~37. 
(A:~89.5\,\% primordial optically-thick, C:~0.0\,\% primordial ultra-settled, B:~0.0\,\% disc-less, D:~7.0\,\% inside-out clearing, E:~0.0\,\% homogeneous draining)}
\label{Tr37}
\end{figure*}

\begin{figure*}
\centering
\includegraphics[width=\textwidth]{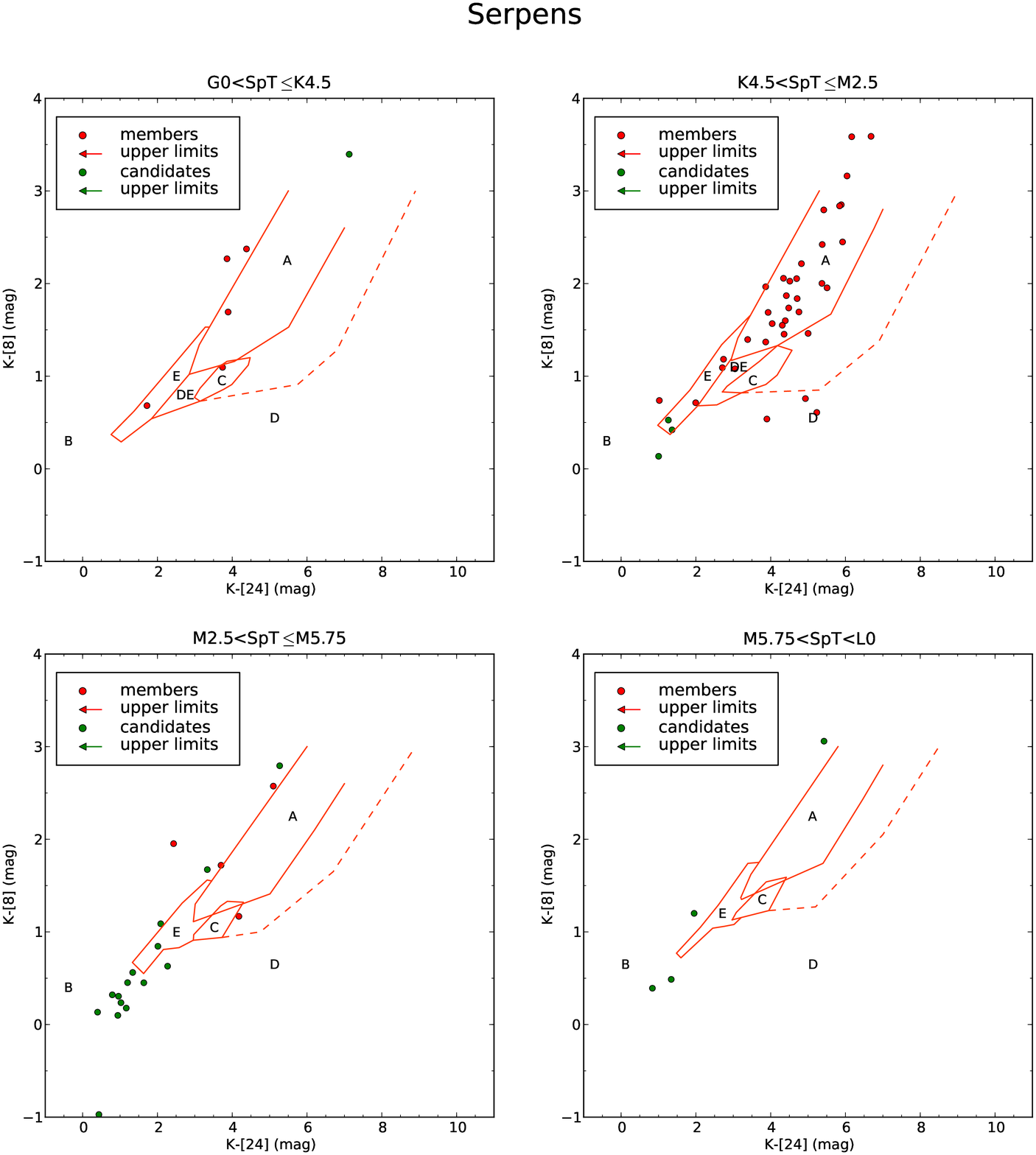} 
\caption{Disc evolution diagnostic diagram (dashed lines represent objects with envelopes) applied to the YSOs in the 2-6 Myr old cluster Serpens. 
(A:~40.6\,\% primordial optically-thick, C:~1.4\,\% primordial ultra-settled, B:~18.8\,\% disc-less, D:~(10.1)\,11.6\% inside-out clearing, E:~(10.1)\,11.6\,\% homogeneous draining) 2 sample points lie outside the limits of this plot, but they are still included in the final statistics.}
\label{Serpens}
\end{figure*}

\begin{figure*}
\centering
\includegraphics[width=\textwidth]{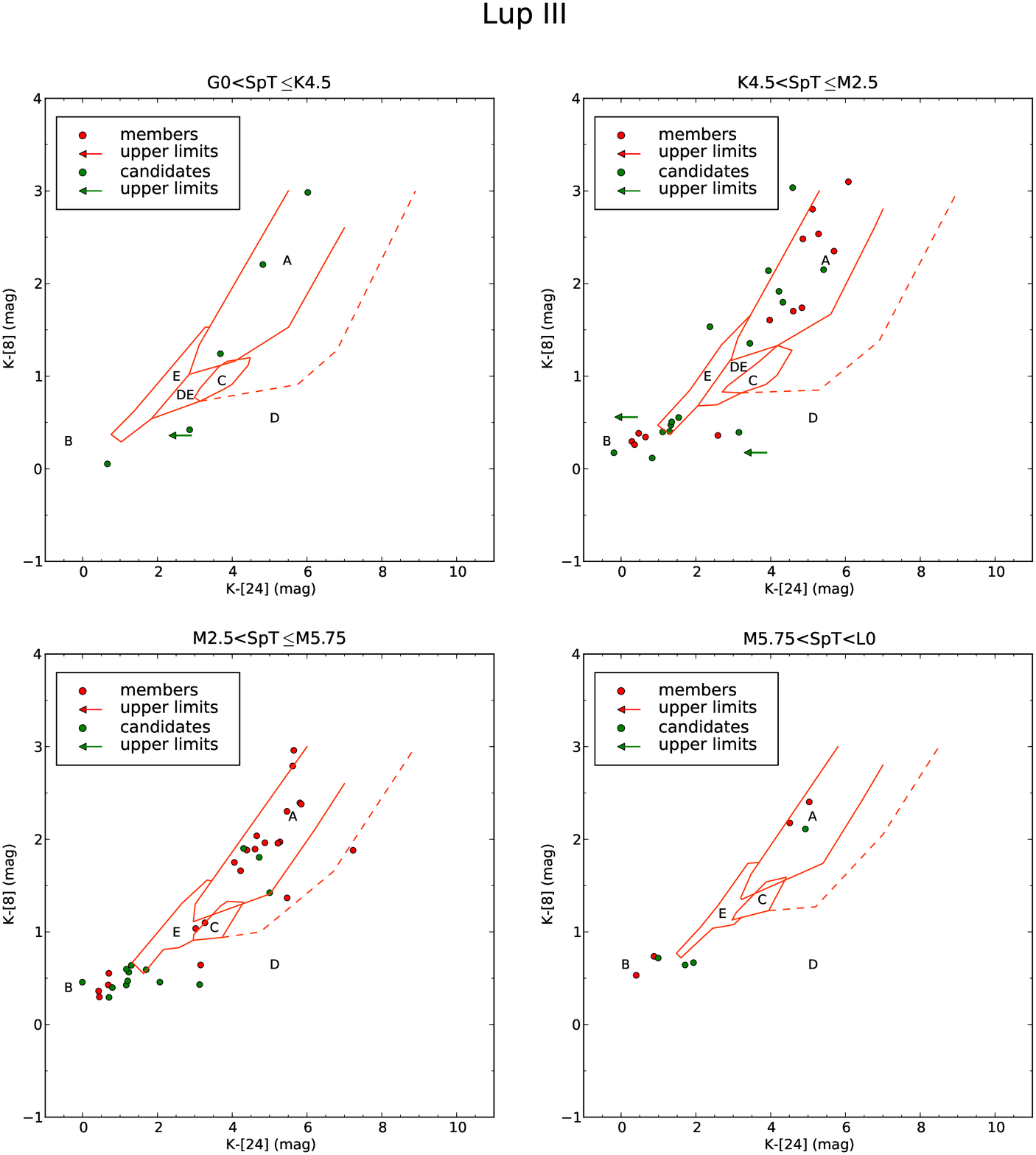} 
\caption{Disc evolution diagnostic diagram (dashed lines represent objects with envelopes) applied to the YSOs in the 2-6 Myr old cluster Lup III. 
(A:~39.0\,\% primordial optically-thick, C:~1.2\,\% primordial ultra-settled, B:~31.7\,\% disc-less, D:~14.6\,\% inside-out clearing, E:~7.3\,\% homogeneous draining) 2 sample points lie outside the limits of this plot, but they are still included in the final statistics.}
\label{LupIII}
\end{figure*}

\begin{figure*}
\centering
\includegraphics[width=\textwidth]{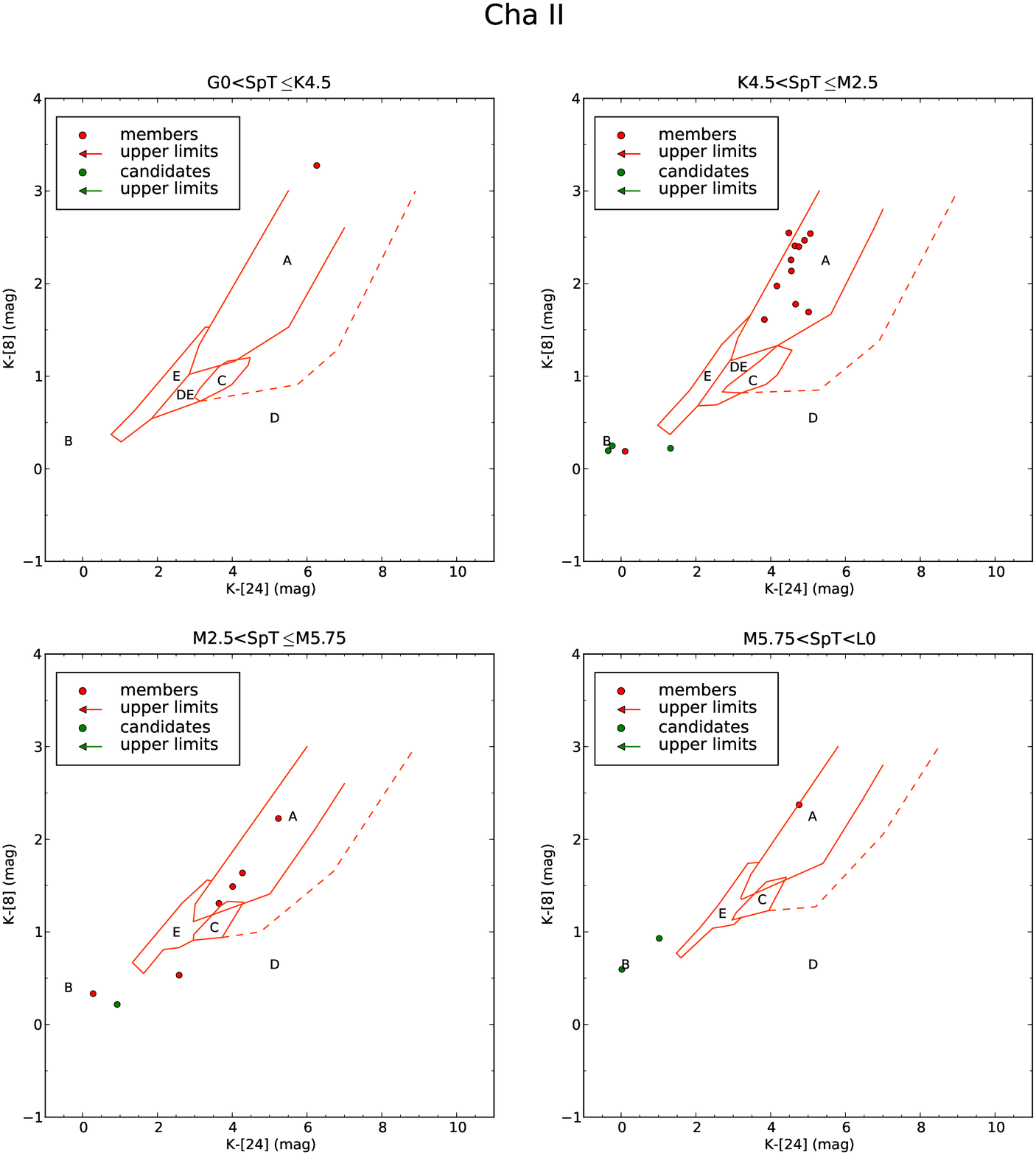} 
\caption{Disc evolution diagnostic diagram (dashed lines represent objects with envelopes) applied to the YSOs in the 4-5 Myr old cluster Cha II. 
(A:~63.0\,\% primordial optically-thick, C:~0.0\,\% primordial ultra-settled, B:~25.9\,\% disc-less, D:~7.4\,\% inside-out clearing, E:~0.0\,\% homogeneous draining) 1 sample point lies outside the limits of this plot, but it is still included in the final statistics.}
\label{ChaII}
\end{figure*}

\begin{figure*}
\centering
\includegraphics[width=\textwidth]{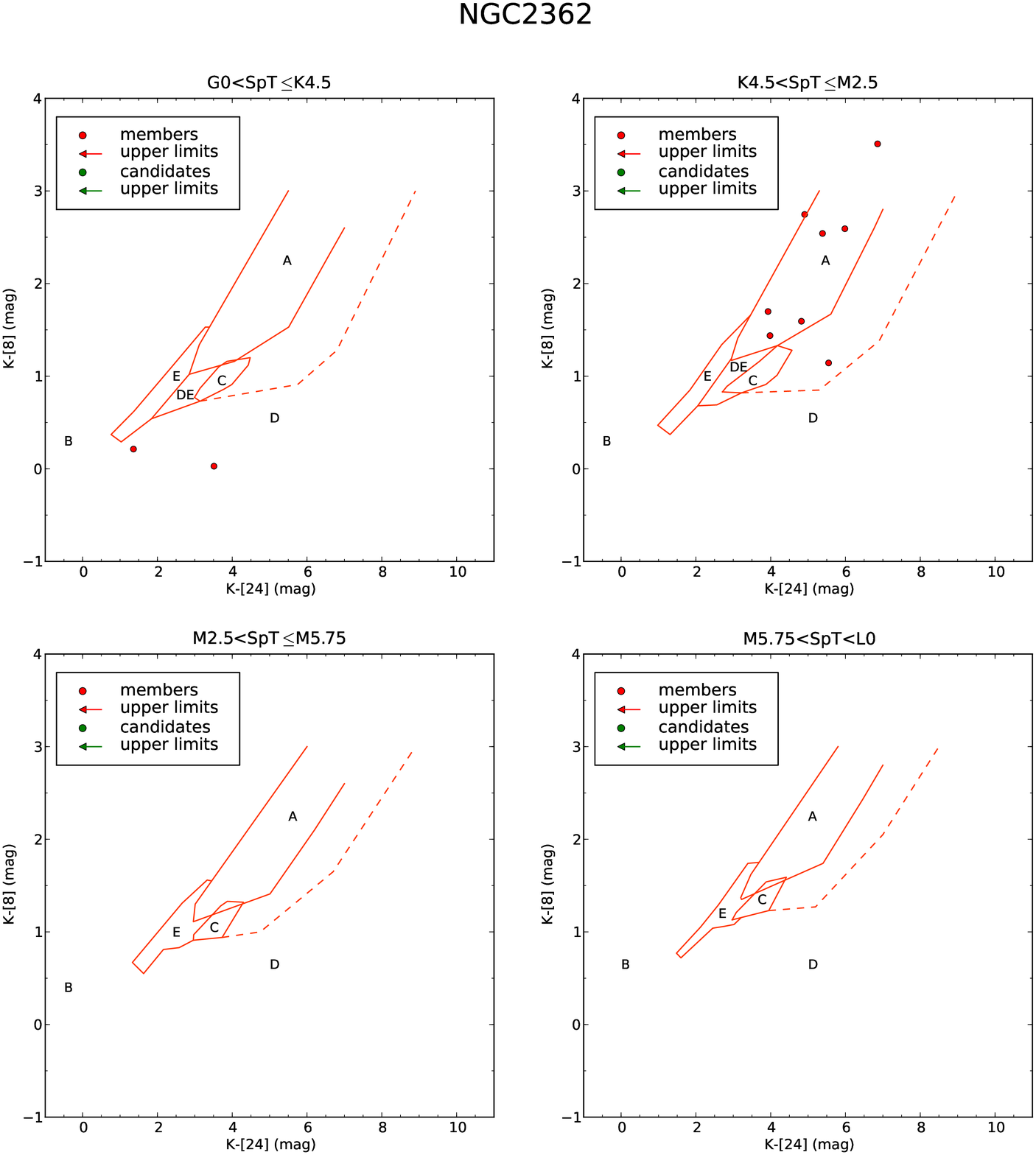} 
\caption{Disc evolution diagnostic diagram (dashed lines represent objects with envelopes) applied to the YSOs in the 5 Myr old cluster NGC2363. 
(A:~60.0\,\% primordial optically-thick, C:~0.0\,\% primordial ultra-settled, B:~0.0\,\% disc-less, D:~30.0\,\% inside-out clearing, E:~0.0\,\% homogeneous draining)}
\label{NGC2362}
\end{figure*}

\begin{figure*}
\centering
\includegraphics[width=\textwidth]{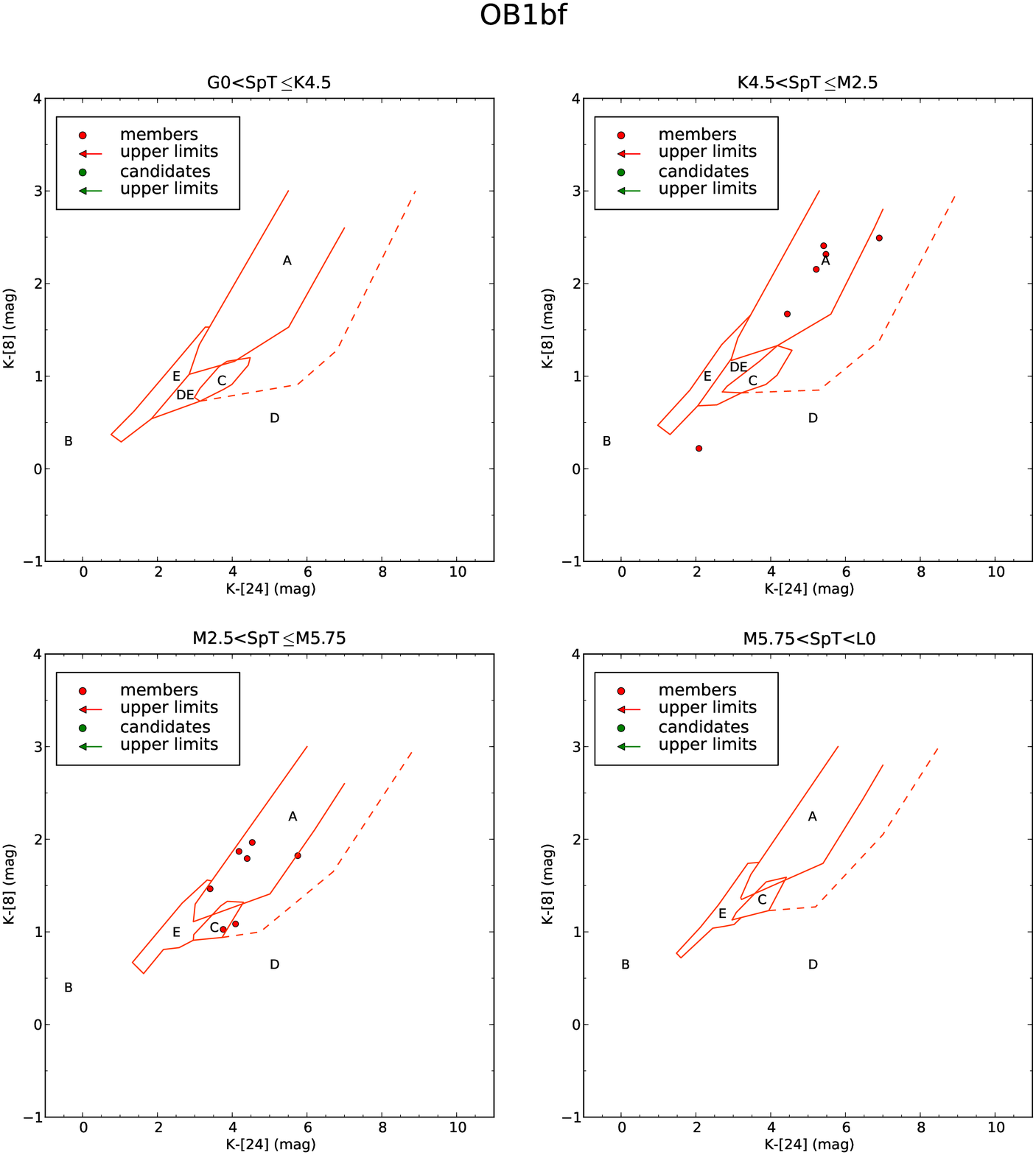} 
\caption{Disc evolution diagnostic diagram (dashed lines represent objects with envelopes) applied to the YSOs in the 5 Myr old cluster OB1bf. 
(A:~61.5\,\% primordial optically-thick, C:~7.7\,\% primordial ultra-settled, B:~0.0\,\% disc-less, D:~30.8\,\% inside-out clearing, E:~0.0\,\% homogeneous draining)}
\label{OB1bf}
\end{figure*}

\begin{figure*}
\centering
\includegraphics[width=\textwidth]{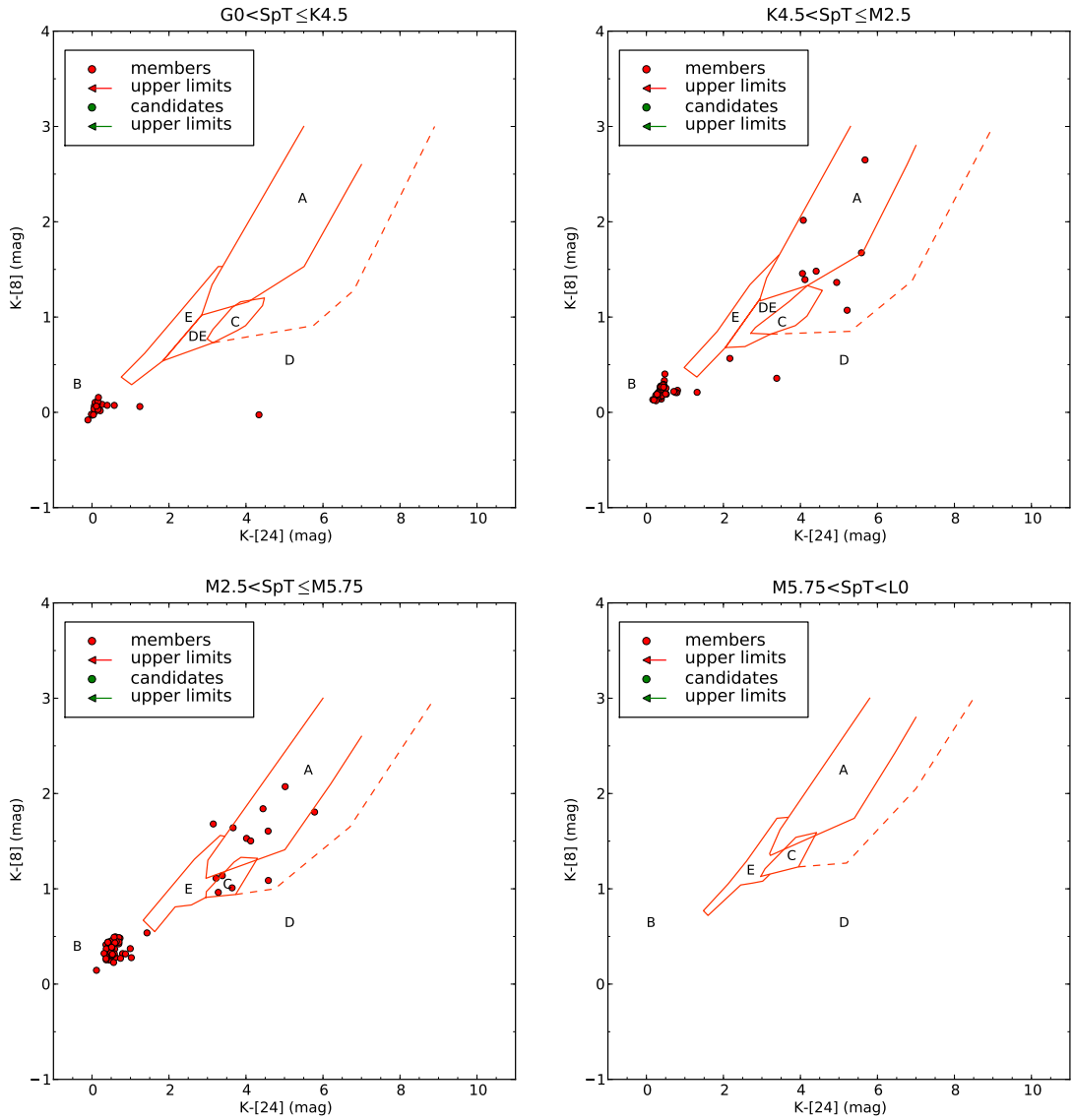} 
\caption{Disc evolution diagnostic diagram (dashed lines represent objects with envelopes) applied to the YSOs in the 5 Myr old cluster Upper Sco. 
(A:~9.0\,\% primordial optically-thick, C:~2.3\,\% primordial ultra-settled, B:~80.5\,\% disc-less, D:~6.8\,\% inside-out clearing, E:~0.8\,\% homogeneous draining)}
\label{UpperSco}
\end{figure*}

\begin{itemize}
\item \emph{Taurus} (Figure~\ref{Taurus-Luhman}, ~\ref{Taurus-Rebull})\\ 
For this 1 Myr \citep{Luhman2004b, Luhman2010, North} old region we have collected data both from \cite{Luhman2010} and \cite{Rebull2010}. The statistics differ when using the two compilations as detailed in what follows. 51.2\,\%  and 41.6\,\% of the Luhman and the Rebull discs, respectively, are still primordial optically-thick. None of the Luhman objects are consistent with being primordial ultra-settled discs, while 0.8\,\% of the Rebull sources fall in this category. 23.2\,\% and 31.5\,\% of objects in the Luhman and Rebull compilations, respectively, have already lost their discs. Currently, 13.7\,\% of the Luhman sources are depleting their discs via the IOC, while between 17.9\,\%  and 19.1\,\%  of the Rebull objects are undergoing that process. Only one disc (0.5\,\%) of the Luhman objects and between 2.3\,\% and 3.5\,\% of the Rebull sources lie in HD region. In the Rebull case we also considered upper and lower limits in the 24\,$\mu m$ band as well as candidates by \cite{Rebull2010}.
\\
\item \emph{NGC2068/71} (Figure~\ref{NGC2068-71})\\
In the 2 Myr \citep{Flaherty2008} old star-forming region NGC2068/71 81.5\,\% of the discs are still classified as primordial optically-thick. 14.8\,\% of the sources are clearing their discs from the inside-out.
\\
\item \emph{NGC2264} (Figure~\ref{NGC2264})\\
25.2\,\% of the discs of the 2 Myr \citep{Teixeira2012} old cluster NGC2264 are still primordial optically-thick. 0.6\,\% are consistent with them being primordial ultra-settled discs. 9.9\,\% have dispersed already and between 56.2\,\% and 57.2\,\% are clearing via the IOC process. Only 0.3\,\% to 1.3\,\% of the discs are found at the edge of the HD regime. We include in this statistics also objects with only upper limit measurements in the 24\,$\mu m$ band. All the objects are classified as candidates by radial velocity studies by \citet{Furesz2006}.
\\
\item \emph{Cha I} (Figure~\ref{ChaI})\\
In Cha I, a 2-3 Myr \citep{Luhman2007,South,Luhman2010} old region, 75.9\,\% of the discs are still primordial optically-thick. One source (3.4\,\%) is clearing its disc from the inside-out, while another disc (3.4\,\%) can be found at the edge of the HD regime.
\\
\ \\
\\
\item \emph{IC348} (Figure~\ref{IC348})\\
23.8\,\% of the objects of the 2-3 Myr \citep{Herbig1998,Muench2003,Luhman2003} old cluster IC348 are still primordial optically-thick. 0.8\,\% are consistent with being primordial ultra-settled discs. 3.8\,\% have dispersed already and between 69.6\,\% to 70.0\,\% are clearing via the IOC process. We include in these statistics also objects with only upper limit measurements in the 24\,$\mu m$ band. 
\\
\item \emph{NGC1333} (Figure~\ref{NGC1333})\\
66.7\,\% of the discs in the less than 3 Myr \citep{Winston2009} old cluster NGC1333 are still primordial optically-thick. 28.2\,\% of sources are depleting their discs via an IOC process.  
\\
\item \emph{Tr~37} (Figure~\ref{Tr37})\\
In the 4 Myr \citep{Sicilia2004,Sicilia2005,Sicilia2006} old star-forming region Tr~37 89.5\,\% of the discs are still primordial optically-thick. None of the objects have primordial ultra-settled discs or have depleted their discs already. 7.0\,\% are currently clearing via the IOC process. For Tr~37 no sources are found in the HD region. Note, that we also included YSO candidates in our statistics.
\\
\item \emph{Serpens} (Figure~\ref{Serpens})\\
For the 2-6 Myr \citep{Oliveira2009} old cluster Serpens we have classified 40.6\,\% of the discs as primordial optically-thick and one disc (1.4\,\%) as primordial ultra-settled. 18.8\,\% of the objects have already lost their discs. Between seven (10.1\,\%) and eight (11.6\,\%) discs are undergoing IOC. Between seven (10.1\,\%) and eight (11.6\,\%) discs lie in the HD regime. Candidates are included in the statistics. 37.5\,\% of the objects in the HD box are classified as candidates by \cite{Oliveira2009}.
\\
\item \emph{Lup III} (Figure~\ref{LupIII})\\
39.0\,\% of the discs in the 2-6 Myr \citep{South,Merin2008} old star-forming cluster Lup III are still primordial optically-thick. One disc (1.2\,\%) has been classified as primordial ultra-settled. 14.6\,\% of discs are undergoing IOC, while 31.7\,\% have already dispersed their discs. We also included YSO candidates and sources for which only upper limits in the 24\,$\mu m$ flux are available in our statistics. In Lup III 7.3\,\% of the discs are found at the photosphere edge of the HD regime. These sources are still classified as candidates by \cite{Merin2008}. 
\\
\item \emph{Cha II} (Figure~\ref{ChaII})\\
In the 4 -5 Myr \citep{Spezzi2008} old star-forming region Cha II 63\,\% of the discs have been classified as primordial optically-thick, while 25.9\,\% have already depleted their discs. 7.4\,\% of the objects are clearing their disc from the inside-out. For ChaII no discs are found in the HD region. We included candidates in our statistics.
\\
\item \emph{NGC2362} (Figure~\ref{NGC2362})\\
The 5 Myr \citep{Mayne2008} old cluster NGC2362 has few spectroscopically-classified objects and/\,or few detected sources in the 24\,$\mu m$ band. 60.0\,\% of the sources are still primordial optically-thick while none are classified as primordial ultra-settled discs or have lost their discs already. 30.0\,\% are clearing their discs via an IOC process.  
\\
\item \emph{OB1bf} (Figure~\ref{OB1bf})\\
61.5\,\% of the discs in the 5 Myr \citep{Hernandez2006} old star-forming region OB1bf are still primordial optically-thick discs and one disc (7.7\,\%) is classified as primordial ultra-settled. 30.8\,\% of the sources are clearing via the IOC process but no sources have lost their discs so far.  
\\
\item \emph{Upper Sco} (Figure~\ref{UpperSco})\\
The 5 Myr \citep{Hernandez2007a} old Upper Sco region seems to be more evolved than most of the other star-forming regions considered in this work. Only 9.0\,\% and 2.3\,\% of the discs, respectively, are still primordial optically-thick or classified as primordial ultra-settled. 80.5\,\% have already lost their discs and 6.8\,\% are clearing their discs via the IOC process. Only one disc (0.8\,\%) is found at the edge of the HD regime.
\\
\item $\eta$ \emph{Cha} (Figure~\ref{etaCha})\\
For the 6 Myr \citep{Mamajek1999} old star-forming region $\eta$ Cha 35.7\,\% of the objects have been classified as primordial optically-thick. 57.1\,\% have already depleted their discs, while one disc (7.1\,\%) is undergoing IOC.
\\
\item \emph{25~Ori} (Figure~\ref{25Ori})\\
In the 7 - 10 Myr \citep{Briceno2007} old star-forming region 25~Ori objects have barely been detected in the 24\,$\mu m$ band. 37.5\,\% of the objects have primordial optically-thick discs, whereas one source (12.5\,\%) has cleared its disc already. 37.5\,\% of the objects are clearing their discs from the inside-out.
\end{itemize}

\subsection{Time-scales}

With the sample of YSOs in different star-forming regions becoming larger and larger, it is now possible to estimate the typical time-scales for the disc dispersal phase, even though cluster ages of course always introduce a large uncertainty in the estimates.
In order to estimate the time-scale of the disc dispersal phase on the basis of our sample, we have considered star-forming regions that have more than 10 objects. We combine boxes A and C into a single 'primordial' category and boxes D and E into a single 'evolved discs' category. All objects in box B remain classified as disc-less sources. We count objects to the left of the A regime in the 'primordial' category, if K\,-\,[8]$>$1.5.

The disc-evolution time-scale of a star-forming region can be estimated by $\tau=\mbox{age}\frac{N_{evolved}}{N_{tot}}$, where age is the estimated age of the star-forming region and $N_{tot}$ and $N_{evolved}$ are the numbers of the total and evolved disc populations, respectively. In Table~\ref{time} we list the individual time-scales for the populations of four spectral type intervals ${\large\tau}${\scriptsize (SpT)} as well as their average $\langle{\large\tau}${\scriptsize (SpT)}$\rangle$ and the cluster time-scale $\tau_{c}$. The absolute errors arise from the error propagation of the standard statistical errors (e.g.\,$\sqrt{N}$) and the uncertainty in age. Note that we arbitrarily assign an error of 10\,\% to those cases where no error is given in the literature.

Typical transition time-scales for the considered star-forming regions are of order $10^{5}$\,yrs. IC348, OB1bf, NGC2264 and Lup III seem to show larger transition time-scales of approximately 1.3 Myrs. The average time-scale across all spectral type $\langle\langle{\large\tau}${\scriptsize (SpT)}$\rangle\rangle=6.9\cdot10^{5}$\,yrs is roughly the same as the average cluster time-scale $\langle\tau_{c}\rangle=6.6\cdot10^{5}$\,yrs. This is partially because there is no significant difference for ${\large\tau}${\scriptsize (SpT)} amongst spectral types. 

\begin{figure*}
\centering
\includegraphics[angle=90,width=\textwidth]{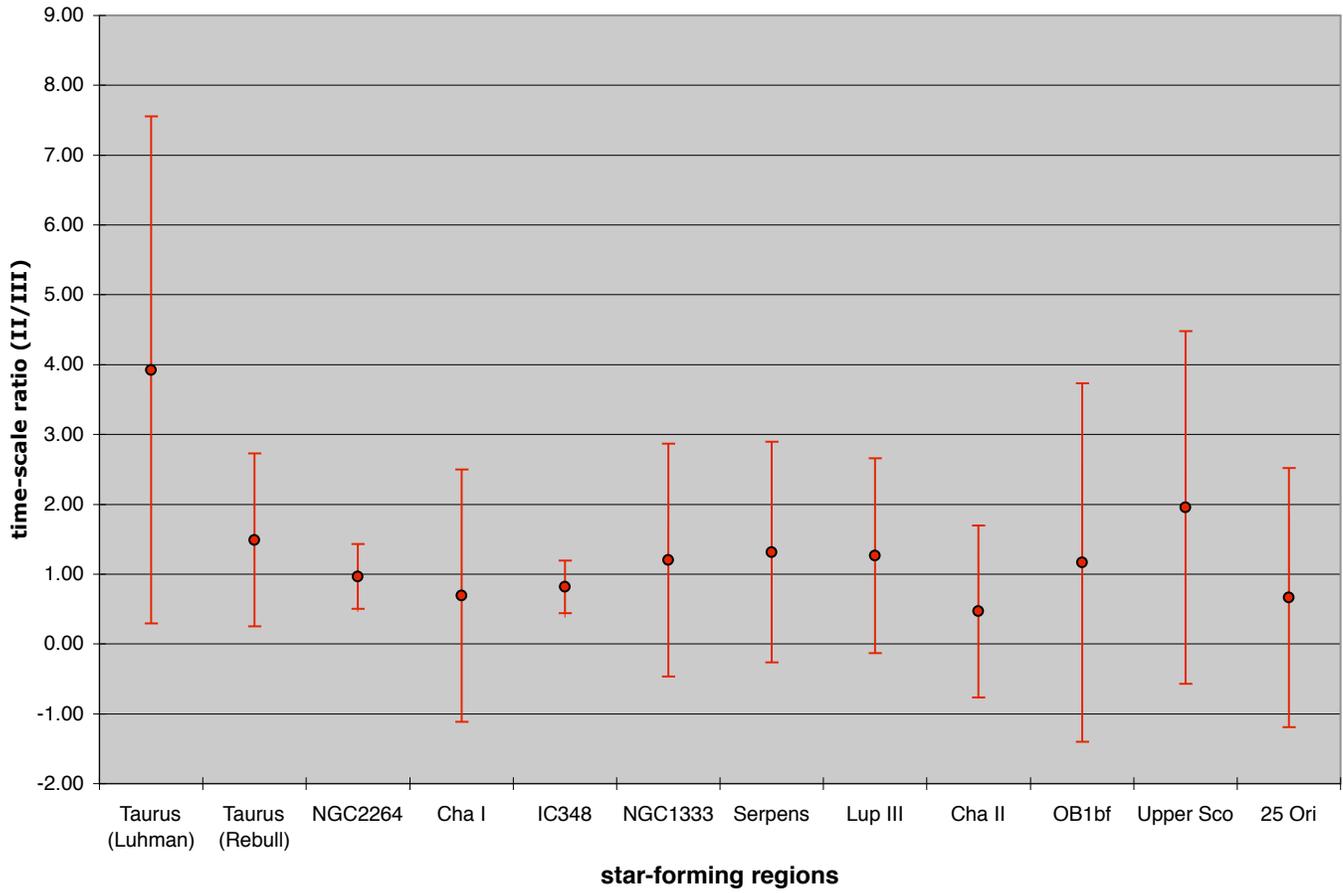} 
\caption{Time-scale ratio ${\large\tau}${\scriptsize (II)}/${\large\tau}${\scriptsize (III)} for K and M stars for the star-forming regions considered in this work with $N_{tot}>10$. Tr37, NGC2068/71 and $\eta$ Cha are not listed, because they lack evolving objects in one or both spectral type intervals.}
\label{timescale23}
\end{figure*}

We further illustrate that disc dispersal time-scales appear to be independent of spectral type. We plot in Figure~\ref{timescale23}, the time-scale ratio ${\large\tau}${\scriptsize (II)}/${\large\tau}${\scriptsize (III)}  (K and M stars) and show that this ratio is consistent with unity. This suggests that there is no significant dependence of the time-scale on stellar mass, as has already been pointed out by \citet{Ercolano2011}, who performed a spatial analysis of the distribution of K and M stars with discs in young star-forming regions and found no significant difference in the distributions. 

\begin{figure*}
\centering
\includegraphics[angle=90,width=\textwidth]{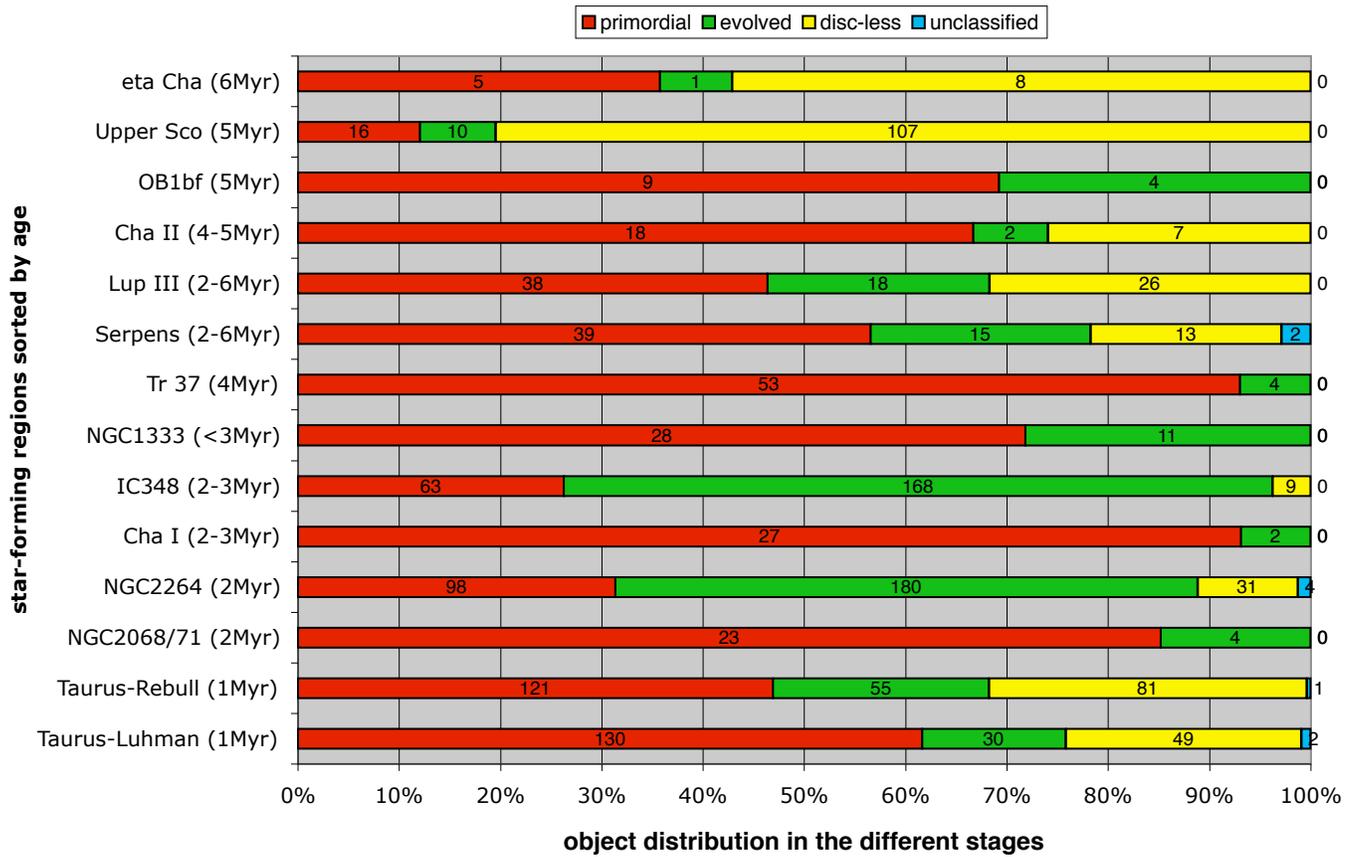} 
\caption{Number of objects in each evolutionary stage (as explained in the legend) as a function of the cluster age.}
\label{timescales}
\end{figure*}

In Figure~\ref{timescales}, we show histograms, correlating the age of the star-forming regions with the number of objects in each evolutionary stage. The expected correlation of percentage of evolved and disc-less objects with age is not clear from the plot. As well as the usual uncertainty in the ages, a number of other factors may contribute to the scatter. Very importantly, identification of class III sources, crucial for the statistics, relies on X-ray observations, to make unbiased classifications. Unfortunately, only 4 out of 15 regions have been sufficiently backed up by X-ray observation. 7 of the regions contain some classified X-ray sources, while the remaining 4 used different criteria for their member classification. 

\section{Summary \& Conclusions}
\label{Conclusion}

We have calculated the SEDs of protoplanetary discs of different spectral types, geometries, settling and inclination. We then considered the evolution of the infrared colours (K\,-\,[8] vs. K\,-\,[24]) of the model disks as they disperse according to different scenarios (homologous depletion, inside-out and outside-in clearing). Based on our models we propose a new diagnostic infrared colour-colour diagram to classify the evolutionary stage of YSOs.
We have applied our infrared colour-colour diagnostic diagram to classify YSOs in 15 nearby star-forming regions and study the evolution of their disc populations.
We estimate time-scales for transition phase of typically a few $10^{5}$\,years independent of stellar mass.

We conclude that, regardless of spectral type, current observations point to fast inside-out clearing as the preferred mode of disc dispersal.

\section*{Acknowledgments}
We thank the referee, Barbara Whitney, for a constructive report that helped us improve the clarity and the solidity of the results presented in our paper.
We would like to thank Dr. Jes\'{u}s Hern\'{a}ndez, who supplied us with spectral types for members of 25Ori and OB1bf.

\bibliographystyle{mn2e}
\bibliography{litnew.bib}
\clearpage

\begin{table*}
 \caption{
Total number of objects and number of objects in each evolutionary regime ( A, B, C, D, E, F*) for the
four spectral type intervals and the 15 star-forming regions considered in this work.}
 \label{statistics}
 \vspace{1cm}
 \rotatebox[origin = c]{90}{  \begin{minipage}[b]{20cm} 
 \begin{tabular}{ccc}
  \begin{tabular*}{9.5cm}{lcrrrrrrr}
  \hline
\multirow{2}{*}{region} & \multirow{2}{*}{SpT} & \multirow{2}{*}{$N_{tot}$} & \multirow{2}{*}{$N_{A}$} & \multirow{2}{*}{$N_{B}$} & \multirow{2}{*}{$N_{C}$} & \multirow{2}{*}{$N_{D}$} & \multirow{2}{*}{$N_{E}$} & \multirow{2}{*}{$N_{F}$}\\
&&&&&&&&\\
  \hline
Taurus$^{L}$& I &8&6&1&-&1&-&-\\
Taurus$^{L}$& II &102&52&25&-&23&-&4\\
Taurus$^{L}$& III &71&34&18&-&4&-&15\\
Taurus$^{L}$& IV &28&16&5&-&1&1&5\\
  \hline  
Taurus$^{R}$& I &46&7&23&1&(11)13&(1)3&1\\
Taurus$^{R}$& II &90&53&18&-&(15)16&(0)1&3\\
Taurus$^{R}$& III &67&25&26&1&6&2&8\\
Taurus$^{R}$& IV &54&22&14&-&14&3&1\\
  \hline  
NGC2068/71& I &9&6&-&-&2&-&1\\
NGC2068/71& II &17&15&-&-&2&-&-\\
NGC2068/71& III &1&1&-&-&-&-&-\\
NGC2068/71& IV &-&-&-&-&-&-&-\\
  \hline 
NGC2264&I&106&20&18&0&(63)65&(0)2&3\\
NGC2264&II&134&41&9&0&(72)73&(1)2&10\\
NGC2264&III&70&17&4&2&40&0&7\\
NGC2264&IV&3&1&0&0&1&0&1\\  
  \hline 
Cha I& I &2&1&-&-&-&-&1\\
Cha I& II &13&12&-&-&1&-&-\\
Cha I& III &9&6&-&-&-&1&2\\
Cha I& IV &5&3&-&-&-&-&2\\
  \hline 
IC348& I &21&1&6&-&(13)14&(0)1&-\\
IC348& II &68&22&2&-&41&-&3\\
IC348& III &138&32&1&2&102&0&1\\
IC348& IV &13&2&-&-&11&-&-\\
  \hline
NGC1333& I &1&1&-&-&-&-&-\\
NGC1333& II &15&10&-&-&5&-&-\\
NGC1333& III &18&11&-&-&5&-&2\\
NGC1333& IV &5&4&-&-&1&-&-\\
 \hline 
Tr~37& I &7&6&-&-&-&-&1\\
Tr~37& II &49&44&-&-&4&-&1\\
Tr~37& III &1&1&-&-&-&-&-\\
Tr~37& IV &-&-&-&-&-&-&-\\
  \hline   
 \end{tabular*}
 &&
 \begin{tabular*}{9.8cm}{lcrrrrrrr}
  \hline
\multirow{2}{*}{region} & \multirow{2}{*}{SpT} & \multirow{2}{*}{$N_{tot}$} & \multirow{2}{*}{$N_{A}$} & \multirow{2}{*}{$N_{B}$} & \multirow{2}{*}{$N_{C}$} & \multirow{2}{*}{$N_{D}$} & \multirow{2}{*}{$N_{E}$} & \multirow{2}{*}{$N_{F}$}\\
&&&&&&&&\\
  \hline 
Serpens& I &7&3&-&1&-&1&2\\
Serpens& II &38&25&2&-&(4)5&(5)6&1\\
Serpens& III &20&-&9&-&3&1&7\\
Serpens& IV &4&-&2&-&-&-&2\\
  \hline 
Lup III& I &6&3&1&-&2&-&-\\
Lup III& II &30&12&8&-&3&4&3\\
Lup III& III &38&14&14&1&5&2&2\\
Lup III& IV &8&3&3&-&2&-&-\\
   \hline 
Cha II& I &1&1&-&-&-&-&-\\
Cha II& II &15&10&3&-&1&-&1\\
Cha II& III &7&4&2&-&1&-&-\\
Cha II& IV &4&2&2&-&-&-&-\\
   \hline 
NGC2362& I &2&-&-&-&2&-&-\\
NGC2362& II &8&6&-&-&1&-&1\\
NGC2362& III &-&-&-&-&-&-&-\\
NGC2362& IV &-&-&-&-&-&-&-\\
  \hline 
OB1bf& I &-&-&-&-&-&-&-\\
OB1bf& II &6&4&-&-&2&-&-\\
OB1bf& III &7&4&-&1&2&-&-\\
OB1bf& IV &-&-&-&-&-&-&-\\
  \hline 
Upper Sco& I &20&-&18&-&2&-&-\\
Upper Sco& II &52&6&41&-&5&-&-\\
Upper Sco& III &61&6&48&3&2&1&1\\
Upper Sco& IV &-&-&-&-&-&-&-\\
  \hline 
$\eta$ Cha& I &-&-&-&-&-&-&-\\
$\eta$ Cha& II &5&1&4&-&-&-&-\\
$\eta$ Cha& III &9&4&4&-&1&-&-\\
$\eta$ Cha& IV &-&-&-&-&-&-&-\\
  \hline 
25 Ori& I &1&1&-&-&-&-&-\\
25 Ori& II &3&1&1&-&1&-&-\\
25 Ori& III &4&1&-&-&2&-&1\\
25 Ori& IV &-&-&-&-&-&-&-\\
  \hline
 \end{tabular*}
 \end{tabular}
 \medskip
\begin{minipage}[b]{\textwidth}
\ \\
*~~unclassified objects\\
()~stand for minimum count, this is to account for objects that fall in \\
the overlap region between boxes D and E, as discussed in Section~\ref{Discs with finite thickness}.\\
\\
L: Data from \citet{Luhman2010}\\
R: Data from \citet{Rebull2010}\\
\begin{eqnarray}
\mbox{I}&\in&[G0; K4.5]\nonumber\\
\mbox{II}&\in&]K4.5; M2.5]\nonumber\\
\mbox{III}&\in&]M2.5; M5.75]\nonumber\\
\mbox{IV}&\in&]M5.75; L0[\nonumber
\end{eqnarray}
\end{minipage}
 \end{minipage}}
 \end{table*}
 \clearpage
 
\begin{table*}
 \caption{The evolutionary time-scale calculated as $\tau=\mbox{age}\frac{N_{evolved}}{N_{tot}}$. We have excluded clusters with $N_{tot}\leq10$.}
 \label{time}
 \vspace{1cm}
 \rotatebox[origin = c]{90}{  \begin{minipage}[b]{20cm} 
 \begin{tabular}{ccc}
  \begin{tabular*}{9.5cm}{llccc}
  \hline
\multirow{2}{*}{cluster}	&	\multirow{2}{*}{SpT}	&	\multirow{2}{*}{$\tau_{SpT}$ [Myr]}			&	\multirow{2}{*}{$<\tau_{SpT}>$ [Myr]}			&	\multirow{2}{*}{$\tau_{c}$	 [Myr]}		\\
&&&&\\
\hline
Taurus$^{L}$	&	all	&-&	0.12	$\pm$	0.10	&	0.14	$\pm$	0.05	\\
Taurus$^{L}$	&	I	&	0.13	$\pm$	0.18	&-&-\\
Taurus$^{L}$	&	II	&	0.22	$\pm$	0.09	&-&-\\
Taurus$^{L}$	&	III	&	0.06	$\pm$	0.04	&-&-\\
Taurus$^{L}$	&	IV	&	0.07	$\pm$	0.07	&-&-\\
\hline
Taurus$^{R}$	&	all	&-&	0.23	$\pm$	0.11	&	0.21	$\pm$	0.06	\\
Taurus$^{R}$	&	I	&	0.30	$\pm$	0.16	&-&-\\
Taurus$^{R}$	&	II	&	0.18	$\pm$	0.08	&-&-\\
Taurus$^{R}$	&	III	&	0.12	$\pm$	0.07	&-&-\\
Taurus$^{R}$	&	IV	&	0.31	$\pm$	0.15	&-&-\\
\hline
NGC2068/71	&	all	&-&	0.34	$\pm$	0.38	&	0.30	$\pm$	0.23	\\
NGC2068/71	&	I	&	0.44	$\pm$	0.51	&-&-\\
NGC2068/71	&	II	&	0.24	$\pm$	0.25	&-&-\\
NGC2068/71	&	III	&-&-&-\\
NGC2068/71	&	IV	&-&-&-\\
\hline
NGC2264	&	all	&-&	1.04	$\pm$	0.57	&	1.15	$\pm$	0.27	\\
NGC2264	&	I	&	1.23	$\pm$	0.39	&-&-\\
NGC2264	&	II	&	1.10	$\pm$	0.33	&-&-\\
NGC2264	&	III	&	1.14	$\pm$	0.43	&-&-\\
NGC2264	&	IV	&	0.67	$\pm$	1.12	&-&-\\
\hline
Cha I	&	all	&-&	0.24	$\pm$	0.36	&	0.17	$\pm$	0.19	\\
Cha I	&	I	&-&-&-\\
Cha I	&	II	&	0.19	$\pm$	0.28	&-&-\\
Cha I	&	III	&	0.28	$\pm$	0.43	&-&-\\
Cha I	&	IV	&-&-&-\\
\hline
IC348	&	all	&-&	1.78	$\pm$	1.05	&	1.75	$\pm$	0.60	\\
IC348	&	I	&	1.67	$\pm$	1.14	&-&-\\
IC348	&	II	&	1.51	$\pm$	0.72	&-&-\\
IC348	&	III	&	1.85	$\pm$	0.71	&-&-\\
IC348	&	IV	&	2.12	$\pm$	1.65	&-&-\\
\hline
NGC1333	&	all	&-&	0.81	$\pm$	0.80	&	0.85	$\pm$	0.48	\\
NGC1333	&	I	&-&-&-\\
NGC1333	&	II	&	1.00	$\pm$	0.81	&-&-\\
NGC1333	&	III	&	0.83	$\pm$	0.65	&-&-\\
NGC1333	&	IV	&	0.60	$\pm$	0.93	&-&-\\
\hline
 \end{tabular*}
 &&
   \begin{tabular*}{9.5cm}{llccc}
\hline
\multirow{2}{*}{cluster}	&	\multirow{2}{*}{SpT}	&	\multirow{2}{*}{$\tau_{SpT}$ [Myr]}			&	\multirow{2}{*}{$<\tau_{SpT}>$ [Myr]}			&	\multirow{2}{*}{$\tau_{c}$	 [Myr]}		\\
&&&&\\
\hline
Tr~37	&	all	&-&	0.33	$\pm$	0.24	&	0.28	$\pm$	0.21	\\
Tr~37	&	I	&-&-&-\\
Tr~37	&	II	&	0.33	$\pm$	0.24	&-&-\\
Tr~37	&	III	&-&-&-\\
Tr~37	&	IV	&-&-&-\\
\hline
Serpens	&	all	&-&	0.81	$\pm$	1.03	&	0.87	$\pm$	0.76	\\
Serpens	&	I	&	0.57	$\pm$	1.07	&-&-\\
Serpens	&	II	&	1.05	$\pm$	1.03	&-&-\\
Serpens	&	III	&	0.80	$\pm$	0.98	&-&-\\
Serpens	&	IV	&-&-&-\\
\hline
Lup III	&	all	&-&	1.00	$\pm$	1.37	&	0.88	$\pm$	0.74	\\
Lup III	&	I	&	1.33	$\pm$	2.15	&-&-\\
Lup III	&	II	&	0.93	$\pm$	0.99	&-&-\\
Lup III	&	III	&	0.74	$\pm$	0.77	&-&-\\
Lup III	&	IV	&	1.00	$\pm$	1.56	&-&-\\
\hline
Cha II	&	all	&-&	0.47	$\pm$	0.68	&	0.33	$\pm$	0.34	\\
Cha II	&	I	&-&-&-\\
Cha II	&	II	&	0.30	$\pm$	0.41	&-&-\\
Cha II	&	III	&	0.64	$\pm$	0.96	&-&-\\
Cha II	&	IV	&-&-&-\\
\hline
OB1bf	&	all	&-&	1.55	$\pm$	1.86	&	1.54	$\pm$	1.35	\\
OB1bf	&	I	&-&-&-\\
OB1bf	&	II	&	1.67	$\pm$	2.03	&-&-\\
OB1bf	&	III	&	1.43	$\pm$	1.69	&-&-\\
OB1bf	&	IV	&-&-&-\\
\hline
Upper Sco	&	all	&-&	0.41	$\pm$	0.35	&	0.38	$\pm$	0.19	\\
Upper Sco	&	I	&	0.50	$\pm$	0.52	&-&-\\
Upper Sco	&	II	&	0.48	$\pm$	0.33	&-&-\\
Upper Sco	&	III	&	0.25	$\pm$	0.20	&-&-\\
Upper Sco	&	IV	&-&-&-\\
\hline
$\eta$ Cha	&	all	&-&	0.67	$\pm$	0.96	&	0.43	$\pm$	0.59	\\
$\eta$ Cha	&	I	&-&-&-\\
$\eta$ Cha	&	II	&-&-&-\\
$\eta$ Cha	&	III	&	0.67	$\pm$	0.96	&-&-\\
$\eta$ Cha	&	IV	&-&-&-\\
\hline
 \end{tabular*}
 \end{tabular}
 \medskip
\begin{minipage}[b]{\textwidth}
\ \\
L: Data from \citet{Luhman2010}\\
R: Data from \citet{Rebull2010}\\
\begin{eqnarray}
\mbox{I}&\in&[G0; K4.5]\nonumber\\
\mbox{II}&\in&]K4.5; M2.5]\nonumber\\
\mbox{III}&\in&]M2.5; M5.75]\nonumber\\
\mbox{IV}&\in&]M5.75; L0[\nonumber
\end{eqnarray}
\end{minipage}
 \end{minipage}}
 \end{table*}
 \clearpage

\clearpage

\clearpage

\clearpage

\clearpage

\clearpage

\clearpage

\clearpage

\clearpage

\clearpage

\clearpage

\clearpage

\clearpage

\clearpage

\clearpage

\begin{figure*}
\centering
\includegraphics[width=\textwidth]{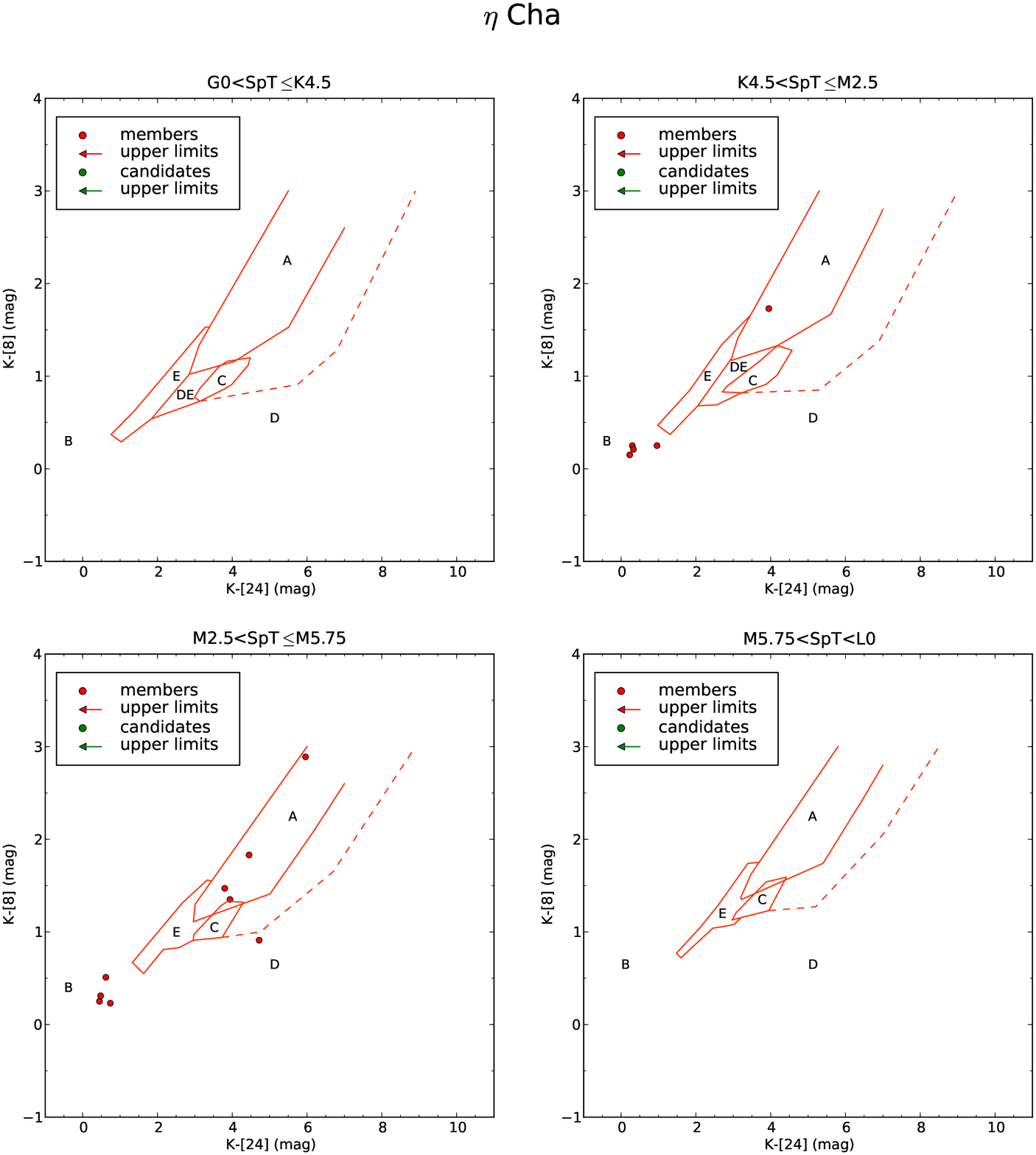} 
\caption{Disc evolution diagnostic diagram (dashed lines represent objects with envelopes) applied to the YSOs in the 6 Myr old cluster $\eta$ Cha. 
(A:~35.7\,\% primordial optically-thick, C:~0.0\,\% primordial ultra-settled, B:~57.1\,\% disc-less, D:~7.1\,\% inside-out clearing, E:~0.0\,\% homogeneous draining)}
\label{etaCha}
\end{figure*}

\clearpage

\begin{figure*}
\centering
\includegraphics[width=\textwidth]{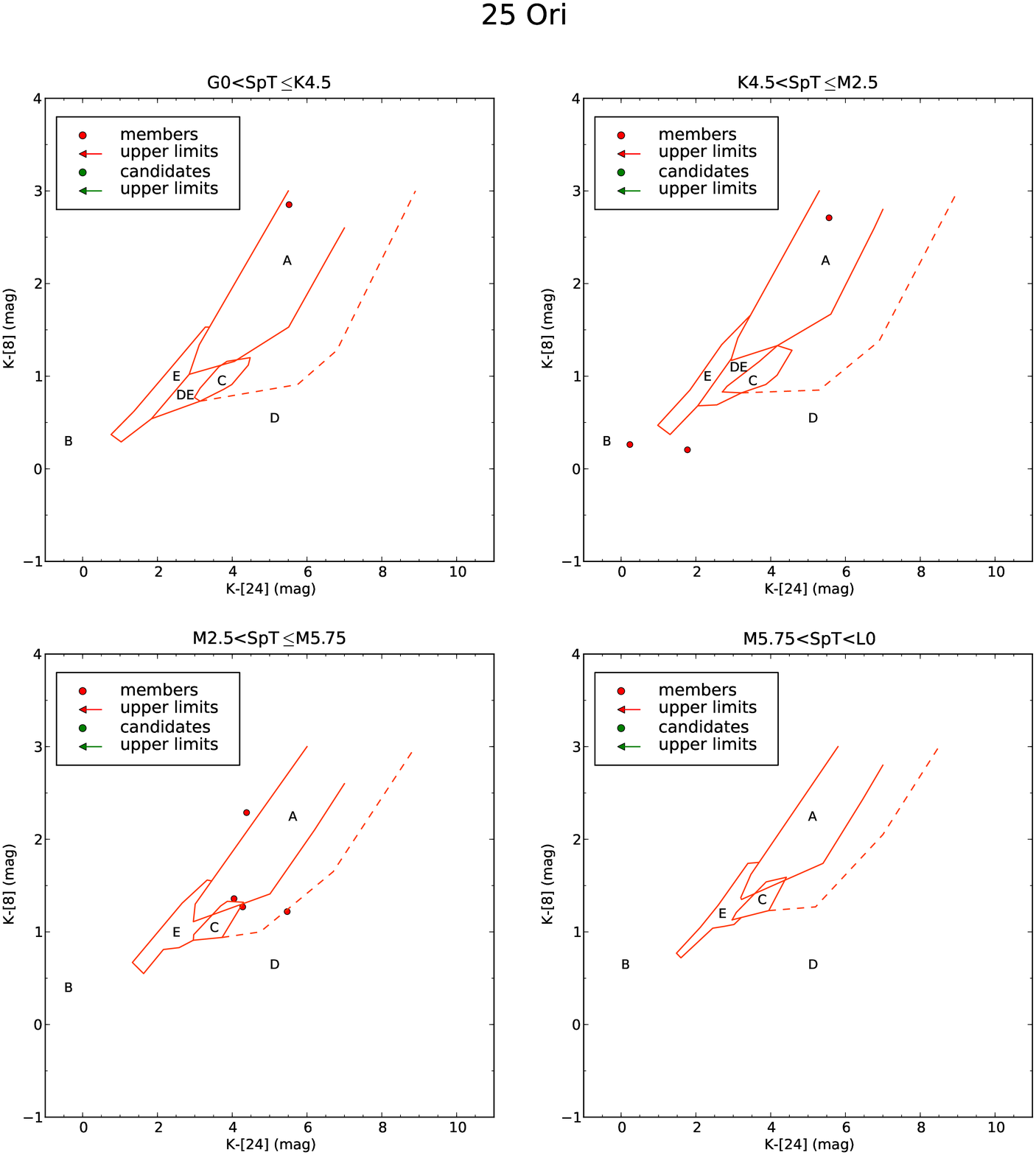} 
\caption{Disc evolution diagnostic diagram (dashed lines represent objects with envelopes) applied to the YSOs in the 7-10 Myr old cluster 25 Ori. 
(A:~37.5\,\% primordial optically-thick, C:~0.0\,\% primordial ultra-settled, B:~12.5\,\% disc-less, D:~37.5\,\% inside-out clearing, E:~0.0\,\% homogeneous draining)}
\label{25Ori}
\end{figure*}

\clearpage

\clearpage

\clearpage

\appendix
\section{}
\label{appendix: boxes}
We list here the boundary points of the evolutionary boxes to allow automatized application to observational data.

\begin{eqnarray}
A^{K\,-\,[24]}_{I}&=& [ 7.,  6.62,  5.51,  4.06,  2.85,  3.12,  5.5 ]\nonumber\\
C^{K\,-\,[24]}_{I}&=&[ 2.99,  3.14,  3.64,  3.86,  4.48,  4.43,  3.99,  3.78,  3.14,  2.99 ]\nonumber\\
E^{K\,-\,[24]}_{I}&=&[ 3.38,  3.28,  2.34,  1.37,  0.76,  1.03,  1.84,  2.85,  1.84,  3.14 ]\nonumber\\
A^{K\,-\,[8]}_{I}&=& [ 2.6,  2.33,  1.53,  1.16,  1.02,  1.34,  3. ]\nonumber\\
C^{K\,-\,[8]}_{I}&=&  [ 0.77,  0.87,  1.1,  1.16,  1.2,  1.12,  0.91,  0.86,  0.73,  0.77 ]\nonumber\\
E^{K\,-\,[8]}_{I}&=&[ 1.53,  1.53,  1.08,  0.62,  0.37,  0.29,  0.54,  1.02,  0.54,  0.73 ]\nonumber
\end{eqnarray}

\begin{eqnarray}
A^{K\,-\,[24]}_{II}&=&[ 7.,  6.77,  5.61,  4.18,  2.95,  2.95,  3.12,  5.3 ]\nonumber\\
C^{K\,-\,[24]}_{II}&=&[ 2.71,  3.22,  3.87,  4.17,  4.57,  4.18,  3.7,  2.83,  2.71 ]\nonumber\\
E^{K\,-\,[24]}_{II}&=&[ 3.44,  2.69,  1.84,  0.98,  1.31,  2.05,  2.95,  2.05,  2.56,  3.22 ]\nonumber\\
A^{K\,-\,[8]}_{II}&=& [ 2.8,  2.6,  1.67,  1.33,  1.17,  1.19,  1.41,  3., ]\nonumber\\
C^{K\,-\,[8]}_{II}&=& [ 0.83,  0.82,  0.91,  1.01,  1.28,  1.33,  1.16,  0.89,  0.83 ]\nonumber\\
E^{K\,-\,[8]}_{II}&=& [ 1.65,  1.34,  0.85,  0.47,  0.37,  0.68,  1.19,  0.68,  0.69,  0.82 ]\nonumber
\end{eqnarray}

\begin{eqnarray}
A^{K\,-\,[24]}_{III}&=&  [ 7.,  6.2,  5.01,  2.96,  3.01,  6. ]\nonumber\\
C^{K\,-\,[24]}_{III}&=& [ 2.96,  2.97,  3.69,  3.87,  4.3,  3.73,  2.96 ]\nonumber\\
E^{K\,-\,[24]}_{III}&=&[ 3.45,  3.33,  2.66,  1.33,  1.63,  2.16,  2.58,  2.96 ]\nonumber\\
A^{K\,-\,[8]}_{III}&=&[ 2.6,  2.1,  1.41,  1.11,  1.3,  3. ]\nonumber\\
C^{K\,-\,[8]}_{III}&=&  [ 0.91,  0.97,  1.28,  1.33,  1.32,  0.94,  0.91 ]\nonumber\\
E^{K\,-\,[8]}_{III}&=&[ 1.55,  1.56,  1.31,  0.67,  0.55,  0.81,  0.83,  0.91 ]\nonumber
\end{eqnarray}

\begin{eqnarray}
A^{K\,-\,[24]}_{IV}&=& [ 7.,  6.46,  5.4,  4.11,  3.22,  3.2,  3.47,  5.8 ]\nonumber\\
C^{K\,-\,[24]}_{IV}&=&[ 2.97,  3.95,  4.42,  3.88,  3.08,  2.97 ]\nonumber\\
E^{K\,-\,[24]}_{IV}&=&[ 3.67,  3.39,  2.6,  2.12,  1.48,  1.6,  2.45,  2.76,  3.02,  3.2 ]\nonumber\\
A^{K\,-\,[8]}_{IV}&=&  [ 2.8,  2.43,  1.74,  1.51,  1.35,  1.37,  1.62,  3. ]\nonumber\\
C^{K\,-\,[8]}_{IV}&=& [ 1.13,  1.23,  1.59,  1.54,  1.21,  1.13]\nonumber\\
E^{K\,-\,[8]}_{IV}&=& [ 1.75,  1.74,  1.29,  1.05,  0.77,  0.72,  1.04,  1.06,  1.08,  1.15]\nonumber
\end{eqnarray}
\\\ \\
From the IOC runs for finite-thickness discs, we can conclude further that all objects found in the D box with colours bluer in K\,-\,[24] than 4 can be considered as evolved ultra-settled objects. Objects redder in K\,-\,[24] than 5 represent the evolved discs which are either settled or mixed and/\,or flared.
\newpage
\vspace*{0.3cm}
The boundaries for primordial discs with envelopes.
\vspace*{0.73cm}
\begin{eqnarray}
\hspace*{2cm}ENV^{K\,-\,[24]}_{I}&=& [3.14,  5.75,  6.8,  8.9]\nonumber\\
\hspace*{2cm}ENV^{K\,-\,[8]}_{I}&=&[0.73,  0.91,  1.28,  3.]\nonumber
\end{eqnarray}
\begin{eqnarray}
\hspace*{2cm}ENV^{K\,-\,[24]}_{II}&=& [3.22,  5.3,  6.9,  9.]\nonumber\\
\hspace*{2cm}ENV^{K\,-\,[8]}_{II}&=&[0.82,  0.85,  1.38, 3.]\nonumber
\end{eqnarray}
\begin{eqnarray}
\hspace*{2cm}ENV^{K\,-\,[24]}_{III}&=&[3.73,  4.75,  6.7,  8.9]\nonumber\\
\hspace*{2cm}ENV^{K\,-\,[8]}_{III}&=&[0.94,  1.0,  1.65,  3.]\nonumber
\end{eqnarray}
\begin{eqnarray}
\hspace*{2cm}ENV^{K\,-\,[24]}_{IV}&=&[3.95,  5.2,  7.,  8.5]\nonumber\\
\hspace*{2cm}ENV^{K\,-\,[8]}_{IV}&=&[1.23,  1.27,  2.05,  3.]\nonumber
\end{eqnarray}
\label{lastpage}

\end{document}